\documentclass[draftclsnofoot, onecolumn, 12pt]{IEEEtran}

\usepackage{amssymb,amsmath,graphicx,paralist,subfigure,pdfpages,cite}
\usepackage{amsmath,amsthm,amssymb,paralist,subfigure,cite,graphicx,algorithm,url,tikz,wasysym,,pdfpages,algorithmic,algorithm,bm}

\newtheorem{thm}{Theorem}

\def\mc{\mathcal}
\def\mb{\mathbf}
\def\mbb{\mathbb}
\def\ra{\rightarrow}
\IEEEoverridecommandlockouts

\newenvironment{noinds_itemize-2}{\begin{list}{}
{\setlength{\rightmargin}{0em}
\setlength{\leftmargin}{0em}
\setlength{\itemsep}{0em}
\setlength{\topsep}{0em}
\setlength{\parsep}{0em}}}{\end{list}}

\def\descriptionnobflabel#1{\hspace\labelsep #1}
\def\descriptionnobf{\list{}{\labelwidth\z@ \itemindent-\leftmargin
 \let\makelabel\descriptionnobflabel}}

\def\s_descriptionlabel#1{\hspace\labelsep \bf #1}

\def\s_descriptionlabel#1{\hspace\labelsep \bf #1}

\newcommand{\bind}[1]{\hspace*{#1}\begin{minipage}[t]{7in}\begin{itemize}}
\newcommand{\eind}{\end{itemize}\end{minipage}\\}

\newcounter{ctr}

\newenvironment{s_itemize-2}{\begin{list}{$\rhd$}
{\setlength{\rightmargin}{0em}
\setlength{\itemsep}{0em}
\setlength{\topsep}{0.25cm}
\setlength{\parsep}{0em}}}{\end{list}}

\newenvironment{itemize-1}{\begin{list}{$\bullet$}
{\setlength{\rightmargin}{0em}
\setlength{\itemsep}{0.5cm}
\setlength{\topsep}{0.25cm}
\setlength{\parsep}{0em}}}{\end{list}}

\newenvironment{itemize-2}{\begin{list}{$\rhd$}
{\setlength{\rightmargin}{0em}
\setlength{\itemsep}{0.25cm}
\setlength{\topsep}{0.25cm}
\setlength{\parsep}{0em}}}{\end{list}}

    {\end{list}}
    {\end{list}}

\newenvironment{itemize-3}{\begin{list}{-}
{\setlength{\rightmargin}{0em}
\setlength{\itemsep}{0.25cm}
\setlength{\topsep}{0.25cm}
\setlength{\parsep}{0em}}}{\end{list}}

\newenvironment{itemize-4}{\begin{list}{\ }
{\setlength{\rightmargin}{0em}
\setlength{\itemsep}{0.25cm}
\setlength{\topsep}{0.25cm}
\setlength{\parsep}{0em}}}{\end{list}}

\newenvironment{short-itemize}{\begin{list}{$\bullet$}
{\setlength{\rightmargin}{0em}
\setlength{\itemsep}{0.10cm}
\setlength{\topsep}{0.10cm}
\setlength{\parsep}{0em}}}{\end{list}}

\font\sf=cmss10
\newcommand{\Nats}{{\hbox{\sf I\kern-.13em\hbox{N}}}}   % Natural numbers
\newcommand{\Reals}{{\hbox{\sf I\kern-.14em\hbox{R}}}}  % Real numbers
\newcommand{\Ints}{{\hbox{\sf Z\kern-.43emZ}}}          % Integers
\newcommand{\CC}{{\hbox{\sf C\kern -.48emC}}}           % Complex numbers
\newcommand{\QQ}{{\hbox{\sf C\kern -.48emQ}}}           % Rational numbers

\newcommand{\lNats}{{\hbox{\sf {\large I\kern-.13em\hbox{N}}}}}   %Natural numbers
\newcommand{\lReals}{{\hbox{\sf {\large I\kern-.14em\hbox{R}}}}}  %Real numbers
\newcommand{\lInts}{{\hbox{\sf {\large Z\kern-.43emZ}}}}          %Integers
\newcommand{\lCC}{{\hbox{\sf {\large C\kern -.48emC}}}}           %Complex numbers
\newcommand{\lQQ}{{\hbox{\sf {\large C\kern -.48emQ}}}}           %Rational numbers

\begin{document}
\title{Localization in internets of mobile agents:\\
	A linear approach}
\author{Sam Safavi,~\emph{Student Member,~IEEE}, Usman A. Khan,~\emph{Senior Member,~IEEE}, Soummya Kar,~\emph{Member,~IEEE},~and Jos\'{e} M. F. Moura,~\emph{Fellow,~IEEE}\thanks{Sam Safavi and Usman A. Khan are with the Department of Electrical and Computer Engineering, Tufts University, 161 College Ave, Medford, MA 02155, USA, {\texttt{\{sam.safavi@,khan@ece.\}tufts.edu}}.
Soummya Kar~and Jos\'{e} M. F. Moura are with the Department of Electrical and Computer Engineering, Carnegie Mellon University, Pittsburgh, PA 15213, USA, {\texttt{\{soummyak,moura\}@andrew.cmu.edu}}.	
This work was supported in part by a National Science Foundation (NSF) CAREER award CCF-1350264 and in part by NSF under grants CCF-1513936 and ECCS-1408222.}
}
\maketitle
\thispagestyle{empty}

\begin{abstract}
Fifth generation~(5G) networks providing much higher bandwidth and faster data rates will allow connecting vast number of static and mobile devices, sensors, agents, users, machines, and vehicles, supporting Internet-of-Things (IoT), real-time dynamic networks of mobile \emph{things}. Positioning and location awareness will become increasingly important, enabling deployment of new services and contributing to significantly improving the overall performance of the 5G~system. Many of the currently talked about solutions to positioning in~5G are centralized, mostly requiring direct line-of-sight (LoS) to deployed access nodes or anchors at the same time, which in turn requires high-density deployments of anchors. But these LoS and centralized positioning solutions may become unwieldy as the number of users and devices continues to grow without limit in sight. As an alternative to the centralized solutions, this paper discusses distributed localization in a 5G enabled IoT environment where many low power devices, users, or agents are to locate themselves without global or LoS access to anchors. Even though positioning is essentially a non-linear problem (solving circle equations by trilateration or triangulation), we discuss a cooperative \textit{linear} distributed iterative solution with only local measurements, communication and computation needed at each agent. Linearity is obtained by reparametrization of the agent location through barycentric coordinate  representations based on local neighborhood geometry that may be computed in terms of certain Cayley-Menger determinants involving relative local inter-agent distance measurements. After a brief introduction to the localization problem, and other available distributed solutions primarily based on directly addressing the non-linear formulation, we present the distributed linear solution for static agent networks and study its convergence, its robustness to noise, and extensions to mobile scenarios, in which agents, users, and (possibly) anchors are dynamic.
\end{abstract}

\section{Introduction}\label{sec1}
With fifth generation~(5G) technologies looming in the horizon, there is a potential for networking vast numbers of heterogeneous devices, the Internet-of-Things~(IoT), not only of static devices, but also of moving objects, users, or vehicles~\cite{koivistocontinuous,witrisal2016high,militano2015device,6815897,Zhang201727,hakkarainen2015high,di2014location}. Location-awareness, providing the physical location of every static or moving object or agent, will enhance the ability to deploy new services and better management of the overall 5G~system. Beyond these, location-aware technologies can also enable a variety of other applications from precision agriculture \cite{ojha2015wireless}, to intruder detection \cite{alrajeh2013intrusion}, health care \cite{4224209}, asset tracking, ocean data acquisition \cite{Bayat:2017}, or emergency services \cite{wymeersch2009cooperative}. For example, location information is essential in providing an effective response in disasters such as fire rescue situations. Other relevant applications include military sensing \cite{rawat2014wireless}, physical security, industrial and manufacturing automation, and robotics \cite{1467103}. In addition, localization is essential in randomly deployed networks, where manual positioning of objects is not practical, and the location of network nodes may change during run-time.

In many scenarios, instrumenting devices with GPS adds to cost and power requirements, reducing the service life of battery-driven devices. Further, GPS receivers are inaccessible in indoor applications, are not effective in harsh environments, and are not sufficiently robust to jamming in military applications \cite{patwari2005locating}. We consider here efficient and low-cost localization algorithms that do not require GPS nor direct access to base station. The solutions to localization, in general, consist of two phases: acquiring measurements and transforming them into coordinate information \cite{gustafsson2005mobile}. In the first phase, nodes\footnote{The terminology node is being used in a generic sense here to denote network entities such as users, agents, anchors. The precise meaning and context will be clear from the mathematical formulations in the following.} collect measurements and exchange information with other network entities, including \emph{neighboring} agents, anchors, or a combination of these. As we discuss in Section~\ref{PF}, the most common measurement techniques include Received Signal Strength (RSS), Time of Arrival (ToA), Time Difference of Arrival (TDoA), and Direction of Arrival (DoA). In the second phase, the information and measurements acquired are aggregated and used as inputs to a localization algorithm~\cite{wymeersch2009cooperative}. The nature of the localization solution, i.e., whether \emph{centralized} or \emph{distributed}, depends on what types of measurements (such as estimates of relative distance between neighboring nodes or distance to possibly distant anchors) are acquired and how the measurements and exchanged information are processed, for instance, whether by a central entity or in a distributed fashion using local computing at the nodes.

Several localization techniques have been proposed in the literature including successive refinements \cite{albowicz2001recursive,savarese2001location}, maximum likelihood estimation \cite{moses2003self,destino2011maximum,kantas2012distributed}, multi-dimensional scaling \cite{shang2004improved}, optimization-based techniques \cite{biswas2006semidefinite,ding2008sensor,cao2005localization}, probabilistic approaches \cite{Ramadurai,peng2007probabilistic},  multilateration \cite{nagpal2003organizing,zhou2012multilateration}, graph theoretical methods \cite{lederer2009connectivity,stanoev2016cooperative}, and ultra-wideband localization \cite{shen2010fundamental1,dardari2009ranging,205552,moura1978passive,zhou2011indoor,bartoletti2011passive,gigl2007analysis}. Other relevant works on the design of localization
algorithms include~\cite{safavi2017localization,sayed2005network,thrun2001robust,langendoen2003distributed,gezici2005localization,conti2012network,blatt2006energy,shen2012network,biswas2011corrective,biswas2013localization,biswas2010wifi,381907,samAsilomar}. In general, the goal is to localize a network of nodes with unknown locations, in the presence of a number of anchors\footnote{Anchors are also referred to as reference nodes, seeds, landmarks, or beacons. In the remaining of this paper, we call the nodes with known locations, anchors, and refer to all other nodes with unknown locations as agents.}. Most of these algorithms consider localization in static networks. However, due to the rapid advances in mobile computing and wireless technologies, mobility is becoming an important area of research in 5G~IoT. Agents can be mobile or mounted on moving entities such as vehicles, robots, or humans. Mobility of nodes increases the capabilities of the network and creates the opportunity to improve their localization. It has been shown that the integration of mobile entities improves coverage, connectivity, and utility of the  network deployment, and provides more data since more measurements can be taken while the agents move~\cite{wang2005sensor,liu2005mobility}. 

In fact, mobility plays a key role in the execution of certain applications such as target tracking \cite{oracevic2014survey}, traffic surveillance \cite{estrin1999next}, and environment monitoring \cite{steere2000research}. For example, in a mobile wireless sensor network monitoring wildfires, mobile sensors can track the fire as it spreads, while staying out of its way~\cite{amundson2009survey}. Mobility also enables nodes to target and track moving objects, e.g., animal tracking for biological research or locating equipment in a warehouse. With advances in wearable technologies, there is a growing interest in localizing IoT objects particularly indoors where GPS-type locationing systems are not available~\cite{borgia2014internet,atzori2010internet,ghayvat2015wsn}, which requires the development of self-localization algorithms that enable each object in the IoT to find its own location by implementing simple object-to-object communication over, e.g., 5G~\cite{Zhang201727,hakkarainen2015high,di2014location}. The mobility of these emergent mobile IoT devices makes the localization problem even more challenging as their locations and available neighborhoods (nearby nodes to implement peer-to-peer communication) keep changing. Therefore, it is important to design fast and accurate localization schemes for mobile networks without compromising the quality of embedded applications that require real-time location information.

As we will discuss in Section~\ref{approaches}, many localization algorithms for mobile networks use sequential Bayesian estimation or sequential Monte Carlo methods. In particular, the Bayesian methods use the recursive Bayes' rule to estimate the likelihood of an agent's location. Due to the non-linear nature of the involved conditional probabilities, the Bayesian solutions are generally intractable and cannot be determined analytically. Solutions involving extended Kalman filters~\cite{roumeliotis2000collective,A1roumeliotis2002distributed,martinelli2005observability,A5martinelli2005multi,kong2006mobile,5152606,cho2011dead,ahmad2013extended,carrillo2013decentralized,A3li2013cooperative}, particle filters~\cite{fox2001particle}, and sequential Monte Carlo methods~\cite{rudafshani2007localization,hu2004localization,dil2006range,yi2007multi,baggio2008monte,stevens2007dual,sheu2010distributed,fox2000probabilistic,thrun2005probabilistic,5432182,	lenser2000sensor} have been proposed to address the associated non-linearities and the intractable computation of the probability distributions. However, these approaches end up being sub-optimal and highly susceptible to the agents' initial guesses of their locations. Despite their implementation simplicity, Monte Carlo or particle filter methods are time-consuming as they need to continue sampling and filtering until enough samples are obtained to represent accurately the posterior distribution of an agent's position~\cite{zhang2008locating}, while requiring a high density of anchors to achieve accurate location estimates.

As an alternative to the traditional non-linear approaches, in this paper we describe a more recent localization framework~\cite{khan2009distributed,khan2015linear,khan2010diland,khanAllerton,khanAllertonMotion}, that, after a suitable reparametrization, reduces to a linear-convex set of iterations implemented in a fully \emph{distributed} fashion and, under broad conditions, is guaranteed to converge to the true agent locations \textit{regardless of their initial conditions}. The linearity in this setup is a consequence of a reparameterization of the nodes' coordinates obtained by exploiting a certain convexity intrinsic to the sensor/agent deployment. In particular, an agent does not update its location as a function of arbitrary neighbors but as a function of a subset of neighbors such that it physically lies inside the convex hull of these neighbors; we call such a set of neighbors, a \textit{triangulation set} of the agent in question\footnote{An avid reader may note that the convex hull associated with this triangulation set constitutes an $m$-simplex in~$\mbb{R}^m$, that is a triangle in a plane or a tetrahedron in~$\mbb{R}^3$.}. Each agent (in the quest of finding its location) updates its location estimate as a function of the location estimates of its neighbors weighted by barycentric coordinates (the reparameterization). The conditions under which the triangulation sets exist, how to find available triangulation sets, and compute the barycentric coordinates have been studied in detail in related work~\cite{khan2009distributed,khan2015linear,khan2010diland,khanAllerton,khanAllertonMotion}, and will be subject of Section~\ref{BC} in the following. 

What is of particular relevance here is that when the set of agents lie inside a triangle (or a square) in~$\mbb{R}^2$ with $3$ (or $4$) anchors at its corners, the aforementioned iterations converge \textit{exponentially fast} to the exact location of each agent regardless of the number of agents\footnote{Technically, the agents need to occupy the space inside the convex hull formed by the anchors in~$\mbb{R}^m$. The smallest non-trivial convex hull in~$\mbb{R}^2$ is a triangle formed by three anchors but lying inside a square or a pentagon with four or five anchors, respectively, is a valid configuration in~$\mbb{R}^2$.}. The resulting framework becomes extremely practical and highly relevant to the IoT settings described earlier as the GPS and line-of-sight to the GPS requirements are replaced with a simple infrastructure requirement of placing, e.g., eight anchors at the eight corners of a big warehouse, office, or a building (effectively a cuboid in~$\mbb{R}^3$). In other words, the global line-of-sight (each agent communicating to the anchor) is substituted with a local line-of-sight where each agent has a communication path that leads to the anchors via neighboring agents. The technical advantages of this framework are also significant as it allows us to treat noise on distance measurements and communication under the purview of stochastic approximation leading to mathematically precise (almost sure) convergence arguments. We will discuss the applicability to imprecise distance measurements and communication in Section~\ref{rnd}.

The next significant advantage of this linear framework is its simplicity in adapting to mobile agents. Traditionally, extending distributed solutions to mobile agents presents many challenges:
\begin{inparaenum}[(i)]
	\item agents may move in and out of the convex hull formed by the anchors;
	\item an agent may not be able to find a triangulation set at all times;
	\item the neighborhood at each agent keeps changing as they move.
\end{inparaenum}
One can imagine an extension of the above linear iterations to this scenario with mobile agents. Here, each agent only updates when its (and other nodes') motion places it inside the convex hull of some other nodes. Location updates thus become opportunistic (time-varying and non-deterministic) as they depend on the availability of nearby agents, which, in turn, is controlled by (possibly) arbitrary and uncoordinated motion models at the agents~\cite{camp2002survey,perkins2003ad,schindelhauer2006mobility,hong1999group}. Although this extension is simple to conceive, finding the conditions under which these iterations converge and the associated rates of convergence are quite non-trivial. We describe the relevant details of this approach in Section~\ref{TRo}.

Another distinguishing feature of this linear framework is that it can be implemented in a completely distributed fashion, i.e., unlike centralized schemes, it does not require any central coordinator to collect and/or process the data. This is because computing barycentric coordinates and the relevant inclusion tests require only local distance information. Distributed algorithms are preferred in 5G, IoT, and other applications where there is no powerful computational center to handle the necessary calculations, or when the large size of the network may lead to a communication bottleneck near the central processor,~\cite{patwari2005locating}.

The rest of this paper is organized as follows. In Section~\ref{tax}, we provide a detailed taxonomy of the existing localization approaches in the literature, and in particular
discuss the emergence of mobility and the role it plays in networked localization. In Section~\ref{PF}, we formulate the localization problem in mobile networks. We briefly discuss Bayesian-based approaches for localization in mobile networks in Section~\ref{approaches}. We then review our linear framework for localization, called DILOC, in static networks in Section~\ref{BC}, and study the extension of DILOC under environmental imperfections in Section~\ref{rnd}. In Section~\ref{TRo}, we show how DILOC extends to mobile networks along with its convergence and noise analysis. We present simulation results in Section~\ref{simul} and finally, Section~\ref{conc} concludes the paper.

\section{Taxonomy of Localization Approaches}\label{tax}
In general, localization problem can be grouped into two main categories: position tracking and global localization. Position tracking, also known as local localization, requires the knowledge of agents' starting locations, whereas global localization refers to the process of determining agents' positions without any prior estimate of their initial locations.
The global localization problem is more difficult,
since the error in the agent's estimate cannot be assumed to
be small,~\cite{thrun2001robust}.
In the remaining of the paper, we only consider the global localization problem. We now present a comprehensive taxonomy of the existing localization algorithms. 

{\bf{Absolute vs. Relative: }}
Absolute localization refers to the process of finding agent locations in a predetermined coordinate system, whereas relative localization refers to such process in a local environment, in which the nodes share a consistent coordinate system. In order to determine the absolute positions, it is necessary to use a small number of anchors that have been deployed into the environment at known locations. These reference nodes define the coordinate system and contribute to the improvement of the estimated locations of the other agents in the network. The location estimates may be relative to a set of anchors at known locations in a local coordinate system, or absolute coordinates may be obtained if the positions of the anchors are known with respect to some global coordinate system, either via GPS or from a system administrator during startup,~\cite{amundson2009survey}. In relative localization, no such reference nodes exist and an arbitrary coordinate system can be chosen. Although increasing the number of anchors in general may lead to more accurate location estimates or may increase the speed of the localization process, the main issue with adding more anchors is that they make the process more expensive, and often become useless after all the agents in the network have been localized.

{\bf{Centralized vs. Distributed:}}
In centralized localization, the locations of all agents are determined by a central coordinator, also referred to as \textit{sink node}. This node first collects measurements (and possibly anchor locations), %via multi-hop routing. It 
and then uses a localization algorithm to determine the locations of all agents in the network and sends the estimated locations back to the corresponding agents. In contrast, such central coordinator does not exist in distributed localization, where each agent infers its location based
on locally collected information.
Despite
higher accuracy in small-sized networks, centralized localization
schemes suffer from scalability issues, and are not feasible in large-scale networks. In addition, comparing to
the distributed algorithms, the centralized schemes are less
reliable and require higher computational complexity due
to, e.g., the accumulated inaccuracies caused by multi-hop
communications over a wireless network.
Distributed algorithms are preferred in applications where there is no powerful computational center to handle the necessary calculations, or 
when the large size of the network may lead to a communication bottleneck near the central processor,~\cite{patwari2005locating}.
Therefore, there has been a growing effort in recent years to develop distributed algorithms with higher accuracies.

{\bf{Conventional vs. Cooperative:}}
In conventional approach, there is no agent-to-agent communication, and therefore agents do not play any role in the localization of other agents in the network. 
For instance, when an agent obtains distance estimates with
respect to three anchors in $\mathbb{R}^2$, it can infer its own location
through \textit{trilateration}, provided the agent knows the
locations of the anchor, see Section~\ref{BC} for more details.
%As an example, consider the traditional \textit{trilateration} for localization in $\mathbb{R}^2$, see Section~\ref{BC} for more details. In this method, each agent first measures its distances to three anchors, and then finds its location as the intersection of three circles whose radii are the distances of the agent to each anchor.
On the other hand, in cooperative approach agents take part in the localization process in a collaborative manner, and are able to exchange information with their neighboring nodes, which may include other agents with unknown locations as well as the anchors. Since anchors are the only reliable source of location information, the former method, also known as \textit{non-cooperative} approach either requires the agents to lie within the communication radius of multiple anchors, which in turn requires a relatively large number of anchors, or to have long-range transmission capabilities in order to make measurements to the anchors. The latter approach removes such restrictions by allowing the inter-agent communications, which in turn increases the accuracy, robustness and the overall performance of the localization process. The reader is referred to Ref.~\cite{wymeersch2009cooperative} for a detailed discussion on the emerge of cooperative localization algorithms.

{\bf{Range-free vs. Range-based:}}
Localization work can also be divided into range-free and range-based methods, based on the measurements used for estimating agent locations. As we discuss in Section~\ref{approaches}, a large number of existing algorithms use distance and/or angle measurements to localize a network of agents, and are therefore referred to as range-based algorithms. Ref.~\cite{Mao20072529} provides an overview of the measurement techniques in agent network localization. Depending on the available hardware, the most common measurement techniques that are used in localization algorithms include Received Signal Strength (RSS), Time of Arrival (ToA), Time Difference of Arrival (TDoA), and Direction of Arrival (DoA),\cite{7869536}. In particular, the distance estimates can be obtained from RSS, ToA, or TDoA measurements; RSS-based localization exploits the relation between power loss and the distance between sender and receiver, and does not require any specialized hardware. However, due to nonuniform signal propagation environments, RSS techniques suffer from low accuracy. 

More reliable distance measurements can be obtained by estimating the propagation time of the wireless signals, which forms the basis of ToA and TDoA measurements. These methods provide high estimation accuracy compared to RSS, but require additional hardware at the agent nodes for a more complex process of timing and synchronization. On the other hand, relative orientations can be determined using AoA measurements, which requires a node to be equipped with directional or multiple antennas. In addition, in order to measure the traveled distance, acceleration, and orientation, an agent may be equipped with an odometer or pedometer, an accelerometer, and an Inertial Measurement Unit (IMU), respectively,~\cite{wymeersch2009cooperative}. On the other hand, range-free algorithms, also known as proximity-based algorithms, use connectivity (topology) information to estimate the locations of the agents. The range-free localization schemes eliminate the need of specialized hardware on each agent, and are therefore less expensive. However, they suffer from lower accuracy compared to the range-based algorithms. Typical range-free localization algorithms include Centroid~\cite{akyildiz2002wireless}, Amorphous~\cite{nagpal2003organizing}, DV-hop~\cite{niculescu2003dv},
%DV-hop follows three steps, namely, location
%broadcast, distance calculation, and location estimation [44].
%In location broadcast, the anchor node broadcasts its location
%information and initializes the hop count to zero among its
%neighbors. The receiver node keeps the minimum hop count
%for each anchor node and disregards the large hop count
%from the same anchor nodes. Then, the receiver increases
%the hop count by one and sends it to the neighbors. Hence,
%each node has a record of the minimum hop count of all
%anchor nodes. In distance calculation, the node calculates the average distance with each anchor node over the hop count
%of all anchor nodes. In location estimation, the blind node
%calculates its location by interlocking the matrixes of anchor
%node location and the matrixes of distance with anchor nodes
%[45]. The disadvantage of hop distance is that it requires a
%uniform distribution of anchor nodes in the whole network
%to achieve high accuracy. Consequently, DV-hop is limited to
%specific applications.
SeRLoc~\cite{lazos2004serloc} and APIT~\cite{he2003range}.

{\bf{Sequential vs. Concurrent:}}
Localization algorithms can also be classified into concurrent methods and sequential methods. 
In concurrent methods, each agent is initially assigned with an estimate of its location coordinates. It then iteratively updates its location estimate using the measurements and the  estimates it acquires from the neighboring nodes. 
The update process continues until the estimates converge to the true coordinates of the agents,~\cite{6857433}.
On the other hand, sequential methods begin with a set of anchors, and compute the locations of the agents in a network, one by one, and in a predetermined sequence. As we explain in Section~\ref{approaches},~\textit{Trilateration} is a common sequential approach, in which each agent computes its location using its distance measurements to $m+1$ anchors in an $m$-dimensional Euclidean space.

{\bf{Static vs. Mobile:}}
Localization literature has mainly focused on static networks where the nodes do not move. This is particularly the case for WSNs, where the problem of locating mobile agents has not been sufficiently addressed. However, due to the rapid advances in mobile computing and wireless communication technologies in recent years, mobility is becoming an important area of research in WSNs. In a particular class of WSNs, namely Mobile Wireless Sensor Networks (MWSNs), agents can be mobile or mounted on moving entities such as vehicles, robots, humans, etc. In MWSNs, mobility of agent nodes increases the capabilities of the network, and creates the opportunity to improve agent localization. It has been shown,~\cite{liu2005mobility}, that the integration of mobile entities into WSNs improves coverage, connectivity, and utility of the agent network deployment, and provides more data since more measurements can be taken while the agents are moving. In addition, mobility plays a key role in the execution of an application. For example, in a MWSN that monitors wildfires, the mobile agents can track the fire as it spreads, and stay out of its way,~\cite{amundson2009survey}. 

Mobility also enables agent nodes to target and track the moving objects. The agent tracking problem is an important aspect of many applications, including the animal tracking, for the purposes of biological research, and logistics, e.g., to report the location of equipments in a warehouse when they are lost and need to be found,~\cite{patwari2005locating}. Another potential application of localization in mobile networks is the \textit{Internet of Things} (IoT), which can be thought of as a massive network of objects such as agents, robots, humans and electronic devices that are connected together and are able to collect and exchange data. With the advancements in wearable technologies, there is a growing interest in localizing IoT objects,~\cite{atzori2010internet,ghayvat2015wsn}, which requires the development of self-localization algorithms that enable each object in the IoT to find its own location by implementing simple object-to-object communications.

Despite all the aforementioned advantages, mobility can make the localization process more challenging; In statically deployed networks, the location of each agent needs to be determined once during initialization, whereas in mobile networks the agents must continuously obtain their locations as they move inside the region of interest. Moreover, mobile nodes require additional power for mobility, and are often equipped with a much larger energy reserve, or have self-charging capability that enables them to plug into the power grid to recharge their batteries,~\cite{amundson2009survey}. It is therefore crucial to design fast and accurate localization schemes for mobile networks without compromising the quality of applications that require wireless communications. 

\section{Preliminaries and Problem Formulation}~\label{PF}
We now formulate the general localization problem in mobile networks. We assume that the agents in the network live in an arbitrary $m$-dimensional, $m\geq 1$, Euclidean space $\mathbb{R}^{m}$, although, for illustration and better visualization of the technical concepts in the conventional 2D or 3D setting, the reader may consider $m=2$ (the plane) or $m=3$. The formalism and results provided below hold for arbitrary $m\geq 1 $ though. In~$\mathbb{R}^{m}, m\geq 1$, consider a network of~$N$ mobile agents collected in a set~$\Omega$ and~$M$ anchors in a set~$\kappa$. The~$N$~agents in~$\Omega$ have unknown locations, while the~$M$ anchors in~$\kappa$ have known locations. Let~$\Theta=\Omega\cup\kappa$ be the set of all nodes, agents and anchors, in the network, and ${\mathcal{N}}_i(k)\subseteq\Theta$ be the set of neighbors\footnote{Note that a neighbor is defined as any node in the set $\Theta$ that lies within the communication radius, $r$, of an agent. Detailed discussion on the notion of neighborhood, quantification of the communication radii, density and distribution of deployment, and related implications can be found in~\cite{khan2009distributed}.} of agent $i\in\Omega$ at time~$k$. Let~$\mb{x}_k^{i*}\in\mbb{R}^m$ be a row vector that denotes the \emph{true location} of a (possibly mobile) node~$i\in\Theta$ at time~$k$, where~$k\geq0$ is the discrete-time index. We describe the evolution of the agent locations as follows:
\begin{equation}\label{evl}
{\bf{x}}_{k+1}^* = f({\bf{x}}_{k}^*,{\bf{v}}_{k}),
\end{equation}
in which~$\mb{x}_k^* \in\mbb{R}^{N\times m}$ is the concatenated state of all agents in the network at time~$k$, the function,~$f$, possibly non-linear, captures the temporal evolution of the locations, and ${\bf{v}}_{k}$ represents the uncertainty in the location evolution at time $k$. Similarly, assume~$\mb{z}_k^i$ to denote the local measurement at agent~$i$ and time~$k$. Let~$\mb{z}_k$ be the collection of all measurements at time~$k$ leading to the following global measurement model:
\begin{equation}\label{ev2}
{\bf{z}}_{k} = g({\bf{x}}_{k}^*,{\bf{n}}_{k}),
\end{equation}
in which ${\bf{n}}_{k}$ captures the measurement noise. The function,~$g$, denotes the measurement technology further explained below. We denote the set of all measurements acquired at time steps,~$1, \ldots,k$, by the set
\begin{equation}\label{BE.1}
{\bf{Z}}_k=\{{\bf{z}}_1,{\bf{z}}_2, \ldots, {\bf{z}}_k\}.
\end{equation}
The localization problem is to estimate the true locations,~$\mb{x}_k^{i*}$, of the mobile agents in the set~$\Omega$ given the measurements in the set~$\mb{Z}$. Since we are seeking distributed solutions, it should also be emphasized that no agent has knowledge of the entire~$\mb{Z}_k$ at any give time; only a subset of~$\mb{Z}_k$ is available at each agent, at time~$k$.

{\bf{Motion model}}:
Regardless of the function~$f$ in Eq.~\eqref{evl}, the motion model can be interpreted as the deviation from the current to the next locations, i.e.,
\begin{eqnarray}\label{Eq1}
\mb{x}_{k}^{i*} = \mb{x}_{k-1}^{i*} + {\widetilde{\mb{x}}_{k}^{i*}},\qquad i\in\Psi,
\end{eqnarray}
in which~${\widetilde{\mb{x}}_{k}^{i*}}$ is the true motion vector at time $k$. We note here that the agents are assumed to move in a bounded region in~$\mbb{R}^m$ and hence~${\widetilde{\mb{x}}_{k}^{i*}}$ cannot take values that drive an agent outside this region.

\section{Localization in Mobile Networks:\\ Bayesian Approaches}\label{approaches}
As we discussed earlier, accurate, distributed localization algorithms are needed for a variety of 5G~IoT and other network applications. In this section, we briefly discuss the basic principles and characteristics of Bayesian methods for solving network localization problems.

{\bf{Baysian Estimation}:}
In general, Bayesian filtering refers to the process of using Bayes' rule to estimate the parameters of a time-varying system, which is indirectly observed through some noisy measurements. In the context of localization, given the history of the measurements up to time~$k$,~${\bf{Z}}_k$ as defined in Eq.~\eqref{BE.1},~Bayesian filtering aims to compute the posterior density,~$p({\bf{x}}_{k}\mid {\bf{Z}}_{k})$, of the agents' locations,~${\bf{x}}_{k}$, i.e., the agent's \textit{belief} about their location at time~$k$.  The goal of localization is to make the belief for each agent as close as possible to the actual distribution of the agent's location, which has a single peak at the true location and is zero elsewhere. Before we explain the Baysian estimation process, let us define the following terms:  The \textit{dynamic model}, $p({\bf{x}}_k\mid {\bf{x}}_{k-1})$, captures the dynamics of the system, and corresponds to the motion model in the context of localization; it describes the agents' locations at time $k$, given that they were previously located at~${\bf{x}}_{k-1}$. The \textit{measurement model},~$p({\bf{z}}_k\mid {\bf{x}}_{k})$, represents the distribution of the measurements given the agents' locations at time $k$. The measurement model captures the error characteristics of the sensors, and describes the likelihood of taking measurement,~${\bf{z}}_k$, given that the agents are located at~${\bf{x}}_{k}$. The Baysian estimation process can then be summarized in the following steps:
\begin{enumerate}
	\item {\bf{Initialization}}: Before the agents start acting (moving) in the environment, they may have initial beliefs about
	where they are. The process then starts with a prior distribution of agent locations,~$p({\bf{x}}_{0})$, which represents the information available on the initial locations of the agents before taking any measurements. For example, in robotic networks such information may be available by providing the robots with the map of the environment. When prior information on agent locations is not available, the prior distribution can be assumed to be uniform.
	\item {\bf{Prediction}}: 
	Suppose the motion model and the belief at time $k-1$, $p({\bf{x}}_{k-1}\mid {\bf{Z}}_{k-1})$, are available.
	Then the predictive distribution of the agents' locations at time $k$
	can be computed as follows:
	\begin{equation}\label{predict}
	p({\bf{x}}_k\mid {\bf{Z}}_{k-1})=\int p({\bf{x}}_k\mid {\bf{x}}_{k-1})p({\bf{x}}_{k-1}\mid {\bf{Z}}_{k-1})d{\bf{x}}_{k-1}.
	\end{equation}
	
	\item {\bf{Update}}: Whenever new
	measurements, ${\bf{z}}_k$, is available,
	the agents incorporate the measurements
	into their beliefs to form new beliefs about their locations, i.e.,
	the predicted estimate gets updated as follows:
	\begin{equation}\label{update}
	p({\bf{x}}_k\mid {\bf{Z}}_{k})= 
	%	{{\alpha}_k}
	\frac{p({\bf{z}}_k\mid {\bf{x}}_{k})p({\bf{x}}_{k}\mid {\bf{Z}}_{k-1})}{p({\bf{z}}_k\mid {\bf{Z}}_{k-1})},
	\end{equation}
	in which the normalization constant,
	%	, ${{\alpha}_k}$, given by
	\begin{equation}
	{p({\bf{z}}_k\mid {\bf{Z}}_{k-1})}=
	%	\frac{1}{{\alpha}_k}=
	\int p({\bf{z}}_k\mid {\bf{x}}_{k})p({\bf{x}}_{k}\mid {\bf{Z}}_{k-1})d{\bf{x}}_{k},
	\end{equation}
	depends on the measurement model and guarantees that the posterior over the entire state space sums up to one.	
\end{enumerate}
In this process, it is standard to assume that
the current locations of the mobile agents, ${\bf{x}}_k$, follow the \textit{Markov}
assumption, which states that 
the agent locations at time $k$ depends only
on the previous location, ${\bf{x}}_{k-1}$, i.e.,
\begin{equation}\label{markov}
p({\bf{x}}_k\mid {\bf{x}}_{0:k-1}, {\bf{Z}}_{k-1})=p({\bf{x}}_k\mid {\bf{x}}_{k-1}).
\end{equation}
In other words, according to the Markovian dynamic model,
the current locations contain all
relevant information from the past.
Otherwise, as the process continues and the number of sensor measurements
increases, the complexity of computing the posterior
distributions grows exponentially over time. 
Under the Markov assumption, the computation cost and memory demand decrease and
the posterior distributions can be efficiently computed  
without losing any information, making the localization process usable in real-time scenarios.

Bayes filters provide a probabilistic
framework for recursive estimation of the agents' locations. However,
their implementation requires specifying
the measurement model, the
dynamic model, and the the posterior distributions. The properties
of the different implementations of Bayes
filters strongly differ in how they represent
probability densities over the state, ${\bf{x}}_{k}$,~\cite{fox2003bayesian}. The reader is referred to~\cite{fox2003bayesian} for a complete survey of the Bayesian filtering techniques. In what follows, we briefly review Kalman filters and particle filters that are the most commonly used variants of Bayesian methods.

{\bf{Kalman Filtering}}:
%\subsection{Kalman Filtering}
When both the dynamic model and the measurement model can be
described using Gaussian density functions, and the initial distribution of the agent locations is
also Gaussian, it is possible to use 
% widely accepted 
Kalman filters, 
%the most widely used
%variant of Bayes filters,
to
derive an exact analytical expression to compute
the posterior distribution of the agent locations.
%Kalman filters are optimal estimators
%The Kalman filter is the closed form solution to the
%Bayesian filtering equations for the filtering model, where the dynamic
%and measurement models are linear Gaussian.
Mathematically, Kalman filters can be used if the dynamic and measurement models can be expressed as follows:
\begin{eqnarray}\label{kf}
{\bf{x}}_{k+1} &=& {\bf{F}}_{k}{\bf{x}}_{k}
%+{\bf{B}}_{k-1}{\bf{u}}_{k-1}
+{\bf{v}}_{k},\nonumber\\
{\bf{z}}_k &=& {\bf{H}}_{k}{\bf{x}}_{k}+{\bf{n}}_{k},
\end{eqnarray}
in which the process noise, %${\bf{v}}_{k}\sim \mathcal{N}({\bf{0}},{\bf{Q}}_{k})$,
\begin{equation*}
{\bf{v}}_{k}\sim \mathcal{N}({\bf{0}},{\bf{Q}}_{k}), 
\end{equation*}
and the measurement noise, %${\bf{n}}_{k}\sim \mathcal{N}({\bf{0}},{\bf{R}}_{k})$, 
\begin{equation*}
{\bf{n}}_{k}\sim \mathcal{N}({\bf{0}},{\bf{R}}_{k}), 
\end{equation*}
are Gaussian with zero mean and covariance matrices of ${\bf{Q}}_{k}$ and ${\bf{R}}_{k}$, respectively, and prior distribution is also Gaussian: %${\bf{x}}_{0}\sim \mathcal{N}(\mathbf{\mu}_0,\mathbf{\Sigma}_0)$. 
\begin{equation*}
{\bf{x}}_{0}\sim \mathcal{N}(\mathbf{\mu}_0,\mathbf{\Sigma}_0).
\end{equation*}
In Eq.~\eqref{kf}, the matrix ${\bf{F}}_{k}$ is the transition matrix and ${\bf{H}}_{k}$ is the measurement matrix.

Due to linear Gaussian model assumptions in Eq.~\eqref{kf}, the posterior
distribution of agent locations is also Gaussian, and can be exactly computed by implementing the prediction, Eq.~\eqref{predict},
and update, Eq.~\eqref{update}, steps using efficient
matrix operations and without any numerical approximations.
The Kalman filter algorithm can be represented with the following recursive algorithm,~\cite{arulampalam2002tutorial}:
\begin{eqnarray}
p({\bf{x}}_{k-1}\mid {\bf{Z}}_{k-1})&=&\mathcal{N}({\bf{m}}_{k-1\mid k-1},{\bf{P}}_{k-1\mid k-1}),\nonumber\\
p({\bf{x}}_{k}\mid {\bf{Z}}_{k-1})&=&\mathcal{N}({\bf{m}}_{k\mid k-1},{\bf{P}}_{k\mid k-1}),\nonumber\\
p({\bf{x}}_{k}\mid {\bf{Z}}_{k})&=&\mathcal{N}({\bf{m}}_{k\mid k},{\bf{P}}_{k\mid k}),\nonumber
\end{eqnarray}
such that
\begin{eqnarray}
{\bf{m}}_{k\mid k-1}&=& {\bf{F}}_{k} {\bf{m}}_{k-1\mid k-1},\nonumber\\
{\bf{P}}_{k\mid k-1}&=& {\bf{Q}}_{k-1}+{\bf{F}}_{k}{\bf{P}}_{k-1\mid k-1}{\bf{F}}^{T}_{k},\nonumber\\
{\bf{m}}_{k\mid k}&=& {\bf{m}}_{k\mid k-1}+{\bf{K}}_k({\bf{z}}_k-{\bf{H}}_k{\bf{m}}_{k\mid k-1}),\nonumber\\
{\bf{P}}_{k\mid k} &=& {\bf{P}}_{k\mid k-1}-{\bf{K}}_k{\bf{H}}_k{\bf{P}}_{k\mid k-1},\nonumber
\end{eqnarray}
where
\begin{eqnarray}
{\bf{S}}_{k}&=& {\bf{H}}_{k} {\bf{P}}_{k\mid k-1}{\bf{H}}^{T}_{k}+{\bf{R}}_{k},\nonumber\\
{\bf{K}}_{k}&=& {\bf{P}}_{k\mid k-1}{\bf{H}}^{T}_{k}
{\bf{S}}^{-1}_{k},\nonumber
\end{eqnarray}
are the covariance of ${\bf{z}}_k-{\bf{H}}_k{\bf{m}}_{k\mid k-1}$, and the Kalman gain, respectively.

%We can express the belief as follows
%\begin{equation}
%p({\bf{x}}_{k}\mid {\bf{Z}}_{k})=\mathcal{N}({\bm{\mu}}_t,{\bf{\Sigma}}_t),
%\end{equation}
%in which, ${\bm{\mu}}_t$ and ${\bf{\Sigma}}_t$ are the distribution's mean and $m \times m$ covariance
%matrix, respectively.
Kalman filters are optimal estimators when the model is linear and Gaussian. They provide efficient,
accurate results if the uncertainty in the location estimates is not too high, i.e., when accurate sensors with high update rates are used,~\cite{fox2003bayesian}.
%Despite Kalman filters' restrictive
%assumptions, practitioners have applied
%them with great success to various tracking
%problems, where the filters yield efficient,
%accurate estimates, even for some
%highly non-linear systems.
However, in many situations such assumptions do not hold, and no analytical solution can be provided.
%the EKF may give a
%large estimation error~\cite{Wang2009}.
In such scenarios, the extended Kalman
filter can be used that approximates the non-linear and non-Gaussian dynamic and measurement models by linearizing the system using first-order Taylor series
expansions.

%Ref.~\cite{panzieri2006multirobot}
{\bf{Particle Filters}}:
%\subsection{Particle Filters}
As discussed earlier, extended Kalman filter results in a Gaussian approximation to
the posterior distribution of the agent locations,
and generates large errors if the true distribution of the belief is not Gaussian. 
In such cases, the \emph{Particle filters} that generalize the traditional Kalman filtering methods to non-linear, non-Gaussian systems can provide more accurate results.
Particle filters, also known as sequential Monte Carlo methods, provide an effective framework to track a variable of interest as it evolves over time,
when the underlying system is non-Gaussian, non-linear, or multidimensional. Unlike other Bayesian filters,
sequential Monte Carlo method is very flexible, easy to
implement and suitable for parallel processing,~\cite{wang2009}.
Since the dynamic and measurement models are often non-linear and non-Gaussian, sequential Monte Carlo methods are in particular widely employed to solve the localization problem in mobile networks.
The particle filter
implementation for the localization problem in mobile wireless sensor networks
% is called
%the Monte Carlo Localization
%(MCL) method, which 
is a recursive Bayesian filter that
constructs a sample-based representation of the entire probability density function, and estimates the
posterior distribution of agents' locations conditioned on their
observations. 
The key
idea is to represent the required posterior probability density function by a
set of random samples or
\textit{particles}, i.e., candidate representations
of agents' coordinates, weighted according to
their likelihood and to compute
estimates based on these samples and weights.
%represent the belief
%represent
%the posterior distribution of agent locations as a %weighted  
%compute the posterior distributions  
%set of samples or
%particles with nonnegative weights.
%, drawn according to the posterior distribution over the observations. 
%
% are a set of simulation-based
%techniques providing a convenient approach 
%in ~\cite{liu2011localization}.
%
%SMC methods are a general class of Monte Carlo methods that sample sequentially from a sequence
%of target probability densities
%
%A series of actions is taken, which modifies the agents' location estimates according to some model. Moreover at certain times an observation arrives that constrains the state of the variable of interest at that time.   
%The SMC scheme estimates the agents' locations in a distributed manner on the basis of the
%connectivity information.
At time $k$, the location estimate of the $i$-th mobile agent, 
%${\bf{x}^j_k}$, 
is represented as a set of $N_s$ samples (particles) , i.e.,
\begin{equation}
S_k^i = \{{\bf{x}}^j_k,w_k^j\},~j=1,\ldots,N_s,
\end{equation}
%the index $j$ denotes the particle and not the agent, 
in which $j$ indicates the particle index, and the weight, $w_k^j$, also referred to as \textit{importance factor}, defines the contribution of the~$j$-th particle to the overall estimate of agent $i$'s location. 
These samples form a discrete representation of the probability density function of the~$i$-th moving agent, and the importance factor is determined by the
likelihood of a sample given agent $i$-th latest observation.
The sequential Monte Carlo localization process for each agent
%Monte Carlo Localization
can then be summarized in the following steps:
%The particle filter algorithm is recursive in nature and operates in two phases: prediction and update.
%
%The localization process in SMC involves three stages, namely, the initial, sample, and filter stages.
\begin{enumerate}
	\item {\bf{Initialization}}:
	At this stage, $N_s$ samples are randomly
	selected from the prior distribution of agent locations,~$p({\bf{x}}_{0})$.
	%	
	%	according to the initial distribution of the system, p(x0).
	%	
	%	 estimates its location by
	%	generating $N_s$ random samples from the region of interest.
	%	Initially $N_s$ particles are randomly distributed
	\item {\bf{Prediction}}: 
	In this step, the effect of the action, from time $k-1$ to $k$, e.g., the movement of the agent according to the dynamic model, is taken into account, and
	$N_s$ new samples are generated to represent the current location estimate.
	%	
	%	If at time $k$, the PDF of the location estimates at the previous instant (time $k-1$) is known, the effect of the actions (from $k-1$ to $k$) should be taken into account in order to update the PDF.
	%	e.g., the motion of the agents,
	%the particles are modified according to the 
	At this point, a random noise is also added to the particles in order to simulate the effect of noise on location estimates. 
	%	In other words, the prediction phase uses a model in order to simulate the effect an action has on the set of particles with the appropriate noise added. 
	%	
	%	In the sample stage, the blind node draws samples in the current
	%	time slot on the basis of the samples from the previous
	%	time slot bounded by a maximum velocity. 
	\item {\bf{
			%measurement update
			Update}}: 	At this stage, sensor measurements are introduced to
	correct the outcome of the prediction step;
	each particle's weight is re-evaluated based on the latest sensory information available,
	%	 i.e., according to the measurement model,
	in order to accurately describe the moving agents' probability density functions.
	%	
	%	In the filter stage, the samples are weighted according
	%	to the anchor node constraint in the current time. Each
	%	valid sample must be within one or two hops of the three
	%	anchor node constraints. Otherwise, the sample is filtered
	%	out.
	\item {\bf{Resampling}}: At this step the particles with very small weights are eliminated, and get replaced with new randomly generated particles so that the number of particles remain constant.
\end{enumerate}
By iteratively applying the above steps, the particle
population eventually converges to the true distribution.
%%
%%%SMC repeats the sampling and filtering process sequentially until sufficient valid samples are obtained.
%%%An estimate of each agent's location is then obtained by the weighted sum of all the particles. 
%Given a particle distribution, three common methods to obtain location estimates are as follows: (i) the weighted mean of the particles; 
%%%\begin{equation}
%%%P_{est}=\sum_{j=1}^p w^j {{\bf{x}}^j},
%%%\end{equation} 
%(ii) the best particle, which corresponds to the one with the maximum weight;
%%%the $P^j$ such that 
%%%\begin{equation}
%%%w^j =\max(w^l),~l \in \{1,\ldots,p\},
%%%\end{equation} 
%and (iii) the robust mean, i.e., the weighted mean in a small window around the best particle. The weighted mean fails in case of multi-modal distributions, whereas the best particle method introduces a discretization error. The robust mean method provides the best result, although it is also the most computationally expensive method,~\cite{rekleitis2004particle}.
Sequential Monte Carlo methods have several key advantages compared with the previous approaches: (i) due to their simplicity, they are is easy to implement; (ii) they are able to represent noise and multi-modal distributions; and (iii) since the posterior distribution can be recursively
computed, it is not required to keep track of the complete
history of the estimates, which in turn reduces the
amount of memory required, and can integrate
observations with higher frequency.
%(3) the particles are universal density approximators. It
%is more accurate than previous approaches.
However, in general sequential Monte Carl methods are very time-consuming 
%Needs sufficient number of particles, depending on
%the size of the environment
because they need to keep sampling and filtering
until enough samples are obtained for representing the
posterior distribution of a moving node's position,~\cite{zhang2008locating}. Moreover, they often require a high density of anchors to achieve accurate location estimates.
%SMC assumes that the time is
%divided into discrete time units and (2) enough samples are
%required at each time slot.
%By adjusting the number of samples
%taken from the belief distribution online, the balance between accuracy and
%computational costs can be adjusted.

In order to avoid the complexity of the above approaches and the susceptibility of their solutions to the initial conditions and the number of available anchors, in the remaining of this paper, we develop a \textit{distributed, linear} framework that exploits convexity to provide a solution to the localization problem both in static and mobile networks. We emphasize that this framework is based on linear iterations that are readily implemented via local measurements and processing at the agents, making it further computationally-efficient compared to the aforementioned centralized and non-linear schemes. Moreover, the proposed solution is distributed, hence reduces the communication and storage overhead associated with, e.g., Bayesian-based solutions.

\section{Localization: A Linear Framework}~\label{BC}
Traditionally, localization has been treated as a non-linear problem that requires either: (i) solving circle equations when the agent-to-anchor distances are given; or, (ii) using law of sines to find the agent-to-anchor distances (and then solving circle equations) when the agent-to-anchor angles are given,~\cite{khan2015linear}. The former approach is called \textit{trilateration}, whereas the latter method is referred to as \textit{triangulation}. The literature on localization is largely based on traditional trilateration and triangulation principles, or, in some cases, a combination of both, see~\cite{khan2015linear} for a historic account; some examples include \cite{niculescu2003ad,liu2006robust,thomas2005revisiting,maxim2008trilateration,zhou2009efficient,khan2009distributed}. 

\begin{figure}[!h]
	\centering
	\subfigure{\includegraphics[width=1.75in]{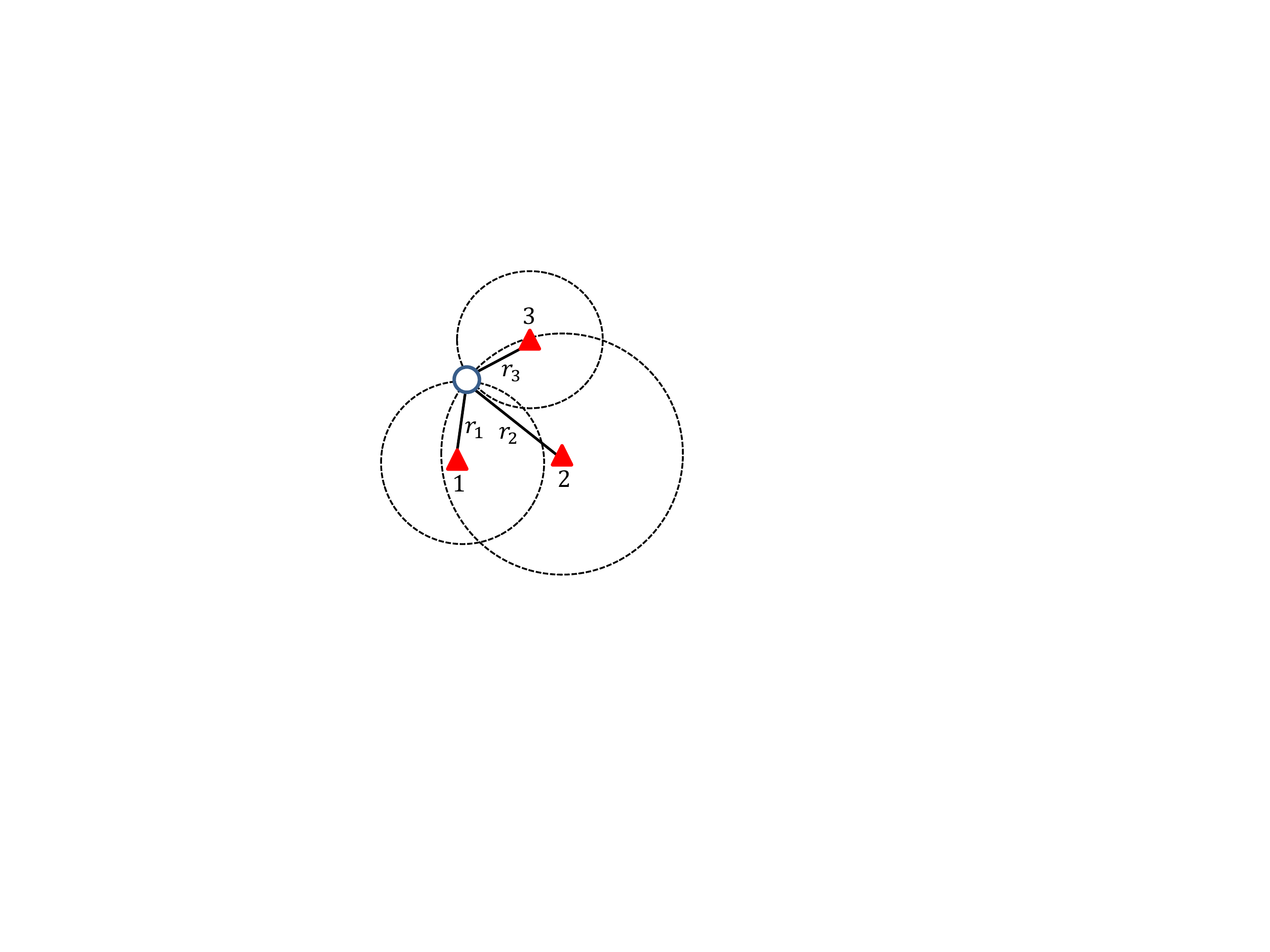}}
	\caption{Traditional trilateration in $\mathbb{R}^2$; the unknown location is at the
		intersection of three circles. Each circle is centered around an anchor, indicated by a red triangle. The radius of each circle is the distance between the agent, represented by a solid circle, and the corresponding anchor.}
	\label{f0}
\end{figure}

In trilateration, the main idea is to first estimate the distances\footnote{As discussed in Section~\ref{PF}, the distances and/or angles of an agent to its nearby nodes can be measured by using RSSI, ToA, TDoA, DoA, or camera-based techniques, \cite{PatwariThesiss,camera}.} between an agent with unknown location and three anchors (in $\mathbb{R}^2$) and then to find the location of the agent by solving three (non-linear) circle equations. As shown in Fig.~\ref{f0}, if the agent with unknown location can measure its distances,~$r_1$,~$r_2$, and $r_3$, to the three nodes with known locations, it can then find its location, $(x,y)\in \mathbb{R}^2$, by solving the following nonlinear circle equations:
\begin{eqnarray}
{(x-x_1)}^2+{(y-y_1)}^2 = r_1^2,\nonumber\\
{(x-x_2)}^2+{(y-y_2)}^2 = r_2^2,\label{eq:nonlinear1}\\
{(x-x_3)}^2+{(y-y_3)}^2 = 
r_3^2,\nonumber
\end{eqnarray}
where $(x_1,y_1)$, $(x_2,y_2)$, $(x_3,y_3)$ represent the coordinates of the three anchors. The placement of the anchors is arbitrary. 

In a large network, like envisioned with IoT, trilateration would need each agent to find its distance to each of three anchors (in two dimensional space) and then solve the nonlinear equations~\eqref{eq:nonlinear1}. Clearly, for large number of agents, this will unduly tax the system resources and be infeasible as it will require \textit{either} placing a large number of anchors so that each agent finds at least three of them \textit{or} long-distance communication and distance/angle estimates from the agents to a small number of possibly far-away anchors. To avoid these difficulties, we discuss a distributed solution to localize a large number of agents in a network. This localization framework can be thought of as a \text{linear} iterative solution to the nonlinear problem posed in terms of the circle equations in~\eqref{eq:nonlinear1}. We call it a framework as many extensions, including, e.g., mobility in the agents and noise on related measurements and communication, rest on this simple framework. 

\subsection{Localization in Static Networks: DILOC} \label{subsec:diloc1}
\begin{figure}[!h]
	\centering
	\subfigure{\includegraphics[width=3.5in]{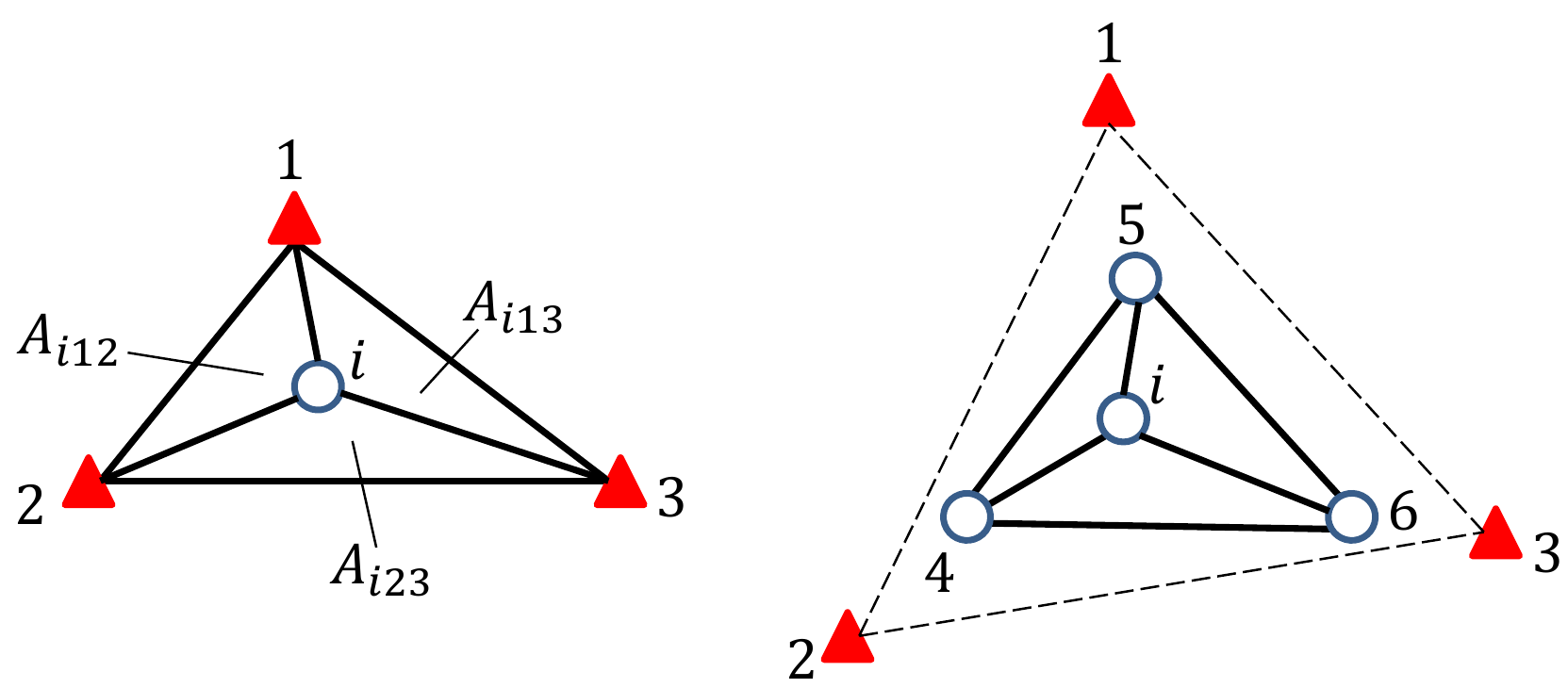}}
	\caption{$\mbb{R}^2$: (Left) Convex-hull inclusion test; (Right) agents, $4$, $5$, and~$6$, form a triangulation set for agent~$i$; all agents are inside the convex hull of the anchors.}
	\label{sim1_5}
\end{figure}

As an alternative to the traditional non-linear approaches, we now present a more recent localization framework, called DILOC,~\cite{khan2009distributed,khan2015linear,khan2010diland,khanAllerton,khanAllertonMotion}, that is based on a reparameterization of the nodes' coordinates by exploiting a certain convexity intrinsic to the agent deployment. Before we proceed, we explicitly denote the true location of the anchors at time~$k$ with $\mathbf{x}_{k}^{i*}\equiv \mb{u}_k^{i\ast}\in\mbb{R}^m, k\in\kappa$, which are known\footnote{Anchors are denoted by red triangles in all of the subsequent figures.}. As briefly mentioned before, in this framework, an agent does not update its location as a function of an arbitrary set of neighbors \textit{but} as a function of a carefully selected subset of neighbors such that it physically lies inside the convex hull of these neighbors, i.e., a triangulation set of the agent in question. This triangulation set is essentially an $m$-simplex in~$\mbb{R}^m$ that is a triangle in a plane or a tetrahedron in~$\mbb{R}^3$. See Fig.~\ref{sim1_5} (Left) where an agent lies inside the convex hull of three anchors and Fig.~\ref{sim1_5} (Right) where an agent lies inside the convex hull of three neighboring agents that are not necessarily anchors. 

In order to describe DILOC, we first assume that all of the nodes are static, i.e., at a any given time, $k$,
\begin{align*}
\mb{x}_k^{i\ast}{}&\equiv\mb{x}^{i\ast},\qquad\forall i\in\Theta,\\
\mb{u}_k^{i\ast}{}&\equiv \mb{u}^{i\ast},\qquad\forall i\in\kappa,
\end{align*}
where $\mb{x}^{i\ast}$'s are unknown and $\mb{u}^{i\ast}$'s are known. Each agent,~$i\in\Omega$, with unknown location in~$\mbb{R}^m$ needs to find a triangulation set, denoted as~$\Theta_i$, of~$m+1$ neighbors (nodes within the communication radius of agent $i$) such that 
\begin{eqnarray}
A_{\mc{C}(\Theta_i)} > 0 \qquad \mbox{and}\qquad 
i\in\mathcal{C}(\Theta_i)\label{eq:convexitycondition1},
\end{eqnarray}
where $\mathcal{C}(\Theta_i)$ denotes the convex hull of the nodes in $\Theta_i$ and $A_{\mc{C}(\Theta_i)}$ denotes the hypervolume of the convex set $\mathcal{C}(\Theta_i)$ in~$\mbb{R}^m$. In \eqref{eq:convexitycondition1}, the first equation states that the nodes in $\Theta_i$ do not lie on a low-dimensional hyper plane in $\mbb{R}^m$, while the second states that agent~$i$ lies strictly inside the convex hull formed by the elements of~$\Theta_i$. In simpler terms, e.g., in~$\mbb{R}^2$, $\mathcal{C}(\Theta_i)$ is a triangle formed by three nodes and $A_{\mathcal{C}(\Theta_i)}$ is its area; additionally, for a set to be a valid triangulation set in~$\mbb{R}^2$, the nodes in the set may not lie on a line in~$\mbb{R}^2$. Each node performs the following test to find a triangulation set. 

{\bf{Convex-hull inclusion test}}: In any arbitrary dimension~$m \geq 1$, at agent~$i$, the following tests determine whether an arbitrary set,~$\overline{\Theta}_i$, of~$m+1$ neighbors is a triangulation set or not:
\begin{eqnarray}\label{convEQ0}
i\in\mathcal{C}(\overline{\Theta}_i),\qquad \mbox{if } \sum_{j\in\Theta_i}A_{\mc{C}(\overline{\Theta}_i\cup\{i\}\setminus j)} = A_{\mc{C}(\overline{\Theta}_i)},\\\label{convEQ1}
i\notin\mathcal{C}(\overline{\Theta}_i),\qquad \mbox{if }\sum_{j\in\Theta_i}A_{\mc{C}(\overline{\Theta}_i\cup\{i\}\setminus j)} > A_{\mc{C}(\overline{\Theta}_i)}.
\end{eqnarray}
Equation~\eqref{convEQ0} states that, when node~$i$ lies inside the convex hull $\mathcal{C}(\overline{\Theta}_i)$, the sum of the areas of the component triangles, $A_{i12}+A_{i23}+A_{i13}$, see the left of Fig.~\ref{sim1_5}, equals the area $A_{123}$ of the triangle formed by the elements of~$\overline{\Theta}_i$. On the other hand,~\eqref{convEQ1} states that, when node~$i$ lies outside the convex hull $\mathcal{C}(\overline{\Theta}_i)$, the sum of the areas of the component triangles is larger than the area of the triangle defined by the elements of~$\overline{\Theta}_i$. Any subset of exactly $m+1$ neighbors that passes the above test becomes a triangulation set,~$\Theta_i$, for agent~$i$. The question is now how to compute the areas of these convex hulls and can these areas be computed by the measurement technologies available at the agents. A neat solution that only requires inter-node distances is provided by the Cayley-Menger determinants, \cite{khan2009distributed,sippl1986cayley}; inter-node distances can be measured (estimated) using the measurement technologies described earlier. 

\textbf{Cayley-Menger determinants: }For any set $\overline{\Theta}_i$ of $m+1$ nodes, in $\mathbb{R}^m$, the Cayley-Menger determinant is the determinant of an~$(m+2)\times(m+2)$ symmetric matrix that uses the pairwise distances of the nodes in~$\overline{\Theta}_i$ to compute (a function of) the hypervolume,~$A_{\overline{\Theta}_i}$, of their convex hull, $\mathcal{C}(\overline{\Theta}_i)$. The Cayley-Menger determinant is given by the following equation:
\begin{equation}\label{cmeq}
A_{\Theta_i}^2=\frac{1}{s_{m+1}}
\begin{vmatrix}
0 & {\bf{1}}_{m+1}^T\\
{\bf{1}}_{m+1} & {\bf{D}}
\end{vmatrix},
\end{equation}
in which~${\mb{1}}_{m+1}$ denotes an $m+1$-dimensional column
vector of $1$'s,~${\mb{D}=\{d_{lj}}^2\},~l,j\in\Theta_i$, is the~$(m+1)\times(m+1)$ Euclidean matrix of squared distances, $d_{lj}$, within the set, $\Theta_i$, and
\begin{equation}\label{cmeq2}
s_m=\frac{2^m{(m!)}^2}{{(-1)}^{m+1}},~m\in\{0,1,2,\ldots \}.
\end{equation}
Although the sequence $s_m$ grows rapidly with~$m$, we are usually locating nodes in~$\mathbb{R}^2$ or in~$\mathbb{R}^3$, for which the second and third coefficients in the above sequence are $-16$ and $288$, respectively. Simply put, given any set of three nodes in~$\mbb{R}^2$, Eq.~\eqref{cmeq} computes the (square of the) area of the triangle (in fact, it is equivalent to Heron's formula on a plane~\cite{heron}) formed by these three nodes. Hence, a convex hull inclusion test can be formulated using these determinants by computing the area of the four underlying triangles. Once a triangulation set~$\Theta_i$ is identified at each agent, (see Remarks in this section for further discussion on the existence and success of finding these sets), we proceed with the reparameterization using the barycentric coordinates as follows:

\textbf{Barycentric coordinates: }
For each agent $i\in\Omega$, we now associate a weight, i.e., its barycentric coordinate to every neighbor $j\in\Theta_i$ in its triangulation set, as follows:
\begin{eqnarray}
p_{ij} &=& \frac{A_{\mc{C}(\Theta_i\cup\{i\}\setminus j)}}{A_{\mc{C}(\Theta_i)}},~~j\in\Theta_i\cap\Omega,\\
b_{ij} &=& \frac{A_{\mc{C}(\Theta_i\cup\{i\}\setminus j)}}{A_{\mc{C}(\Theta_i)}},~~j\in\Theta_i\cap\kappa.
\end{eqnarray}
We note that the barycentric coordinate associated to a non-anchor neighbor in the triangulation set is denoted by a lowercase $p$, while the one associated to an anchor neighbor is denoted by a lowercase $b$. A triangulation set may not contain any anchor in which case all of these weights are designated with lowercase $p$. The reason for splitting the barycentric coordinates into non-anchors and anchors will become apparent later. We note here that the barycentric coordinates\footnote{As a side note, these coordinates are attributed to the work by M{\"o}bius \cite{mobius1827barycentrische}, while a simpler version of the Cayley-Menger determinant is associated to Lagrange, see~\cite{khan2015linear} for a historical account.} are positive (ratio of hypervolumes) and they sum up to one by~\eqref{convEQ0}:
\begin{align}
\sum_{j\in\Theta_i} 
\label{eq:sumtoone-1}
\left(p_{ij}+b_{ij}\right)=\sum_{j\in\Theta_i\cap\Omega} p_{ij}+\sum_{j\in\Theta_i\cap\kappa}b_{ij}=1.
\end{align}

With all the ingredients in place, i.e., the convex hull test, triangulation sets, and the barycentric coordinates, we now present the linear algorithm that exploits all of this convex geometry and the associated coordinate reparameterization.

\textbf{Algorithm: }Before we proceed, recall that the agents are static with true locations,~$\mb{x}^{i*}$, and their location estimate at time~$k$ is denoted by~$\mb{x}_k^i$. We would like to have an algorithm that converges to (or learns) the true locations, i.e.,~$\mb{x}_k^i\ra\mb{x}^{i*}$. To this aim, each agent $i\in\Omega$ with a triangulation set $\Theta_i$ updates its location estimate as follows:
\begin{equation}\label{DILOCeqMain}
\mb{x}^i_{k+1} = \sum_{j\in\Theta_i\cap\Omega}  p_{ij}\mb{x}_k^j + \sum_{j\in\Theta_i\cap\kappa} b_{ij}{\mb{u}^j}^\ast,\qquad i\in\Omega.
\end{equation}
Clearly, when a triangulation set does not contain any anchor ($\Theta_i\cap\kappa=\emptyset$), the second part of the above iteration is empty and the entire update is in terms of the neighboring non-anchor agents (in $\Omega $) that also do not know their locations. Since there are~$N$ agents in the set~$\Omega$, there are a total of~$N$ vector equations~\eqref{DILOCeqMain} that we refer to as barycentric system of equations. 

To get some insight into~\eqref{DILOCeqMain} consider one agent and two anchors on a line, with true positions $c^*, a^*, b^*$, respectively, such that $a^*<c^*<b^*$. One can verify that $c^*\in(a^*,b^*)$ and~\eqref{DILOCeqMain} reduces to
\begin{align*}
c_{k+1}&=\frac{b^*-c^*}{b^*-a^*}a^*+\frac{c^*-a^*}{b^*-a^*}b^* = c^*.
\end{align*}
The barycentric coordinates are a function of the length of the line segments (hyper-volume of a $m=1$-simplex) that are computed with the distance estimates. Since, in this simple example, the agent expresses its location in terms of the two anchors, DILOC provides the correct location in one-step. When the neighbors are non-anchors, $a^*$ and $b^*$ are also unknown and are (naturally) replaced by their estimates, i.e., $a_k$ and $b_k$, and we study how this linear, time-invariant, system of equations evolve over time, and where does it converge (if it does). The analysis for arbitrary~$\mbb{R}^m$ is considered next. 

\textbf{Analysis: }We now write the barycentric system of equations~\eqref{DILOCeqMain} in compact matrix-vector notation by defining
\begin{alignat*}{3}
\mb{x}_{k+1}{}&=\left[\begin{array}{c}
\mb{x}^1_{k+1}\\
\cdots\\
\mb{x}^N_{k+1}
\end{array}\right], \qquad
&&\mb{u}^*&&=\left[\begin{array}{c}
\mb{u}^1\\
\cdots\\
\mb{u}^{M}
\end{array}\right]\\
\mb{P}{}&=\left\{p_{ij}\right\}\in[0,1)^{N\times N},\qquad 
&&\mb{B}&&=\left\{b_{ij}\right\}\in[0,1)^{N\times M},
\end{alignat*}
where $[0,1)^{N\times M}$ denotes an $N\times M$ matrix with elements in the interval $[0,1)$. 
Then, in vector format,~\eqref{DILOCeqMain} becomes
\begin{eqnarray}\label{DILOCeqVect}
\mbox{DILOC:}\qquad\qquad\mb{x}_{k+1} = \mb{P}\mb{x}_k + \mb{B}\mb{u}^*.
\end{eqnarray}
The reason for splitting the barycentric coordinates is now apparent. In \eqref{DILOCeqVect}, the matrices, $\mb{P}$ and $\mb{B}$, collect the barycentric coordinates and the vectors, $\mb{x}_k$ and $\mb{u}^*$, the agent and anchor coordinates, respectively. DILOC's convergence is summarized in the following theorem.
\begin{thm}\label{DILOCthm}
In~$\mbb{R}^m$, assume that the agents lie inside the convex hull of a non-coplanar set of anchors, i.e.,~$\mc{C}(\Omega)\in\mc{C}(\kappa)$ with $A_{\mc{C}(\kappa)}>0$. If each agent successfully finds a triangulation set, then DILOC in~\eqref{DILOCeqVect} converges to the true agent locations. 
\end{thm}

The detailed proof is beyond the scope of this paper and can be found in~\cite{khan2009distributed}. Here we provide a brief overview. First note that Eq.~\eqref{DILOCeqVect} can be further rewritten as
\begin{eqnarray}\label{DILOCeqVect-2}
\left[\begin{array}{c}
\mb{u}^*\\
\mb{x}_{k+1} 
\end{array}
\right]
= \underbrace{\left[\begin{array}{cc}
\mb{I}&\mb{0}\\
\mb{B}& \mb{P}
\end{array}
\right]}_{\triangleq\boldsymbol{\Upsilon}}
\left[\begin{array}{c}
\mb{u}^*\\
\mb{x}_k 
\end{array}\right].
\end{eqnarray}
Intuitively, the system matrix~$\bf{\Upsilon}$ of the above LTI system can be interpreted as a transition probability matrix of a Markov chain because it is non-negative and its rows sum to $1$; recall that the barycentric coordinates are positive and sum to~$1$, see~\eqref{eq:sumtoone-1}. This Markov chain has exactly~$m+1$ absorbing states, because of the identity $m\times m$ block in the upper left corner, while the rest of the states are transient. Under the purview of the convexity condition that all agents lie inside the convex hull of the anchors in Theorem~\eqref{DILOCthm} and that each agent successfully triangulates (the latter is in fact a consequence of the former), one can show that this is an absorbing Markov chain, i.e., all transient states have a path to the absorbing states, resulting into~$\rho(\mb{P})<1$, where~$\rho$ denotes the spectral radius of a matrix. Successive iterations thus result into:
\begin{align*}
\boldsymbol{\Upsilon}^{k+1}=\left[\begin{array}{cc}
\mathbf{I}_{M} & \mathbf{0}\\
\sum_{\ell=0}^{k}\mathbf{P}^{\ell}\mathbf{B} & \mathbf{P}^{k+1} \end{array}\right],
\end{align*}
and in the limit
\begin{align*}
\lim_{k\rightarrow\infty}\boldsymbol{\Upsilon}^{k+1}=\left[\begin{array}{cc}
\mathbf{I}_{M} & \mathbf{0}\\
 \left(\mathbf{I}_{N}-
\mathbf{P}\right)^{-1}\mathbf{B} & \mathbf{0} \end{array}\right],
\end{align*}
from which it follows that
\begin{align}
\label{eq:agentsconverge-1}
\lim_{k\rightarrow\infty}\mb{x}_k{}&=\left(\mathbf{I}_{M}-
\mathbf{P}\right)^{-1}\mathbf{B}\bm{u}^*,
\end{align}
which in fact constitutes the true coordinates of the agents expressed in terms of the barycentric coordinates with respect to the anchors~\cite{khan2009distributed}. Note that the limit in~\eqref{eq:agentsconverge-1} to the true agent locations results regardless of the initial conditions.

\textbf{DILOC Remarks: }It is imperative to discuss the intuition behind some of the arguments made while presenting DILOC and its advantages over traditional setups. 
\begin{enumerate}[(i)]
\item We first discuss the assumption that all agents strictly lie inside the convex hull of the anchors, i.e.,~$\mc{C}(\Omega)\in\mc{C}(\kappa)$. With the help of this assumption, one can show that each agent will successfully find a triangulation set, not necessarily composed by anchors. The exact probability of successfully finding a triangulation set in a small radius around an agent depends on the density of the agents and the probability distribution of the deployment. Details of these arguments can be found in~\cite{khan2009distributed}. 

\item Next, for a non-coplanar set of anchors in~$\mbb{R}^m$, one must have at least~$m+1$ anchors providing a lower bound on the number of anchors. Since we assumed that~$\mc{C}(\Omega)\in\mc{C}(\kappa)$, all the agents must \textit{either }lie inside the triangle formed by three anchors in~$\mbb{R}^2$ \textit{or} inside the tetrahedron formed by four anchors in~$\mbb{R}^3$. We emphasize that this deployment condition to localize an arbitrary number of static agents in~$\mbb{R}^m$ is exactly the same as localizing one static agent in~$\mbb{R}^m$. In this sense, DILOC provides a scale-free solution. 

\item The resulting framework becomes extremely practical and highly relevant to the IoT settings described earlier as the GPS and line-of-sight to the GPS requirements are replaced with a simple infrastructure requirement of placing, e.g., eight anchors at the eight corners of a big warehouse, office, or a building (effectively a cuboid in~$\mbb{R}^3$). In other words, the global line-of-sight (each agent communicating to the anchor) is substituted with a local line-of-sight where each agent has a communication path that leads to the anchors via neighboring agents. 

\item The iterative linear convex DILOC in~\eqref{DILOCeqMain} or~\eqref{DILOCeqVect} contrasts with traditional trilateration. Trilateration requires each agent to solve the nonlinear problem given in~\eqref{eq:nonlinear1} and leads to major drain on the agents' and/or anchors' resources while creating a major communication bottleneck, as the anchors must continuously broadcast their message to \textit{all} the agents, no matter how far they are. In 5G environments, trilateration will require line-of-sight from every agent or user or device to anchors. In comparison, DILOC provides a linear (iterative) solution and only requires local communication, no LoS to anchors but only to neighboring nodes. The power needed by agents to communicate locally to their neighbors in their triangulation set is steerable, in the following sense. At the set-up phase, when agents look to determine who are other nodes in their triangulation set, they can start transmitting with a minimum power level. If at that level, they find $m+1$ other nodes they can transmit at this power level. If they cannot find $m+1$ such nodes, they increase the power level to reach further away nodes. Once they reach $m+1$ nodes there is no need to increase further the transmitting power.

\item There are several issues that have been addressed in the literature. For example, incorporating all available agents in the DILOC update instead of only~$m+1$, incorporating all available anchors instead of the minimum~$m+1$, developing randomized strategies to improve on the worst-case convergence rates, extending the setup to include agents that do not reside inside the anchor convex-hulls, extending to continuous-time algorithms to tackle on frequency-dependent noise, incorporating environmental imperfections, and extending to mobile agents which move arbitrarily in a bounded region. We refer the interested readers to the following relevant body of literature on this topic:~\cite{khan2010diland,khanAllerton,khanAllertonMotion,4663899, kar2010distributed,6857433,hou2013framework,deghat2011distributed,kang2008barycentric}. 
\end{enumerate}
In the next sections, we consider two of the most significant extensions that are built on this framework, i.e., addressing communication noise and imperfect distance measurements in Section~\ref{rnd} and incorporating agent mobility and non-deterministic motion in Section~\ref{TRo}.

\section{DILOC in Random Environments}\label{rnd}
We now consider DILOC under environmental imperfections that are relevant to IoT settings. We make the following assumptions on the types of noises and disturbances.
\begin{noinds_itemize-2}
	\item \textbf{Assumption R0: Data packet drops:} Peer-to-peer communication on ad hoc wireless networks is marred with data packet drops. In order to address such packet drops or communication link failures, we assume that the communication link~$(i\leftarrow j)$ is active with a non-zero probability,~$q_{ij}\in(0,1]$, at each iteration. At time~$k$, we model this link by a binary random variable,~$e_{ij}(k)$, such that
	\begin{equation}\label{rnd1}
	e_{ij}(k)= \left\{
	\begin{array}{ll}
	1,&\mbox{w.p. }q_{ij},\\
	0,&\mbox{w.p. }1-q_{ij},
	\end{array}
	\right.
	\end{equation}
	where~$e_{ij}(k)=1$ shows perfect communication at time~$k$.
	
	\item \textbf{Assumption R1: Communication noise:} In order to address communication noise over peer-to-peer wireless networks, we model the data exchange to be imperfect and corrupted with additive noise. In particular, agent~$i$ receives only a corrupt version,~$\mathbf{y}_{ij}(k)$, of
	node~$j$'s state,~$\mathbf{x}_{j}(k)$, given by
	\begin{equation}\label{rnd2}
	\mathbf{y}_{ij}(k)=\mathbf{x}_{j}(k)+\mathbf{v}_{ij}(k).
	\end{equation}
	The components of the noise vector,~$\mathbf{v}_{ij}(k)$, are independent and zero-mean with finite second moments.
	
	\item \textbf{Assumption R2a: Noisy distance measurements:} The DILOC setup of Section~\ref{BC} assumes perfect knowledge of inter-node distances to compute the barycentric coordinates. However, these distances may not be known perfectly in reality. We now assume that the agents only know distance estimates,~$\widehat{d}_{ij}(k)$, computed from noisy distance measurements at time~$k$, e.g., Received-Signal-Strength, RSS, or Time-of-Arrival, TOA, see~\cite{khan2009distributed,khan2010diland,khan2010hdc,khan2015linear} for further details on this setup. Due to imperfect distances, the system matrices,~$\mathbf{P}$ and~$\mathbf{B}$ from~\eqref{DILOCeqVect}, become a function of both time,~$k$, and distance estimates, denoted by $\widehat{\mathbf{P}}_k(\widehat{\mathbf{d}}_k)$ and ~$\widehat{\mathbf{B}}_k(\widehat{\mathbf{d}}_k)$, where we ignore the parentheses in the sequel to simplify notation. Assuming the distance measurements to be statistically independent over time, we split the system matrices as
	\begin{align}\label{rnd3}
	\widehat{\mathbf{P}}_k&=&\mathbf{P}+\mathbf{S}_\mathbf{P}+ \widetilde{\mathbf{S}}_\mathbf{P}(k)\triangleq\{\widehat{p}_{ij}(k)\},\\
	\widehat{\mathbf{B}}_k&=&\mathbf{B}+\mathbf{S}_\mathbf{B}+
\widetilde{\mathbf{S}}_\mathbf{B}(k)\triangleq\{\widehat{b}_{ij}(k)\},
	\end{align}
	where~$\mathbf{S}_\mathbf{P}$ and~$\mathbf{S}_\mathbf{B}$ are measurement error biases, and
	$\{\widetilde{\mathbf{S}}_\mathbf{P}(k)\}_{k\geq0}$ and
	$\{\widetilde{\mathbf{S}}_\mathbf{B}(k)\}_{k\geq0}$ are
	independent random matrices with zero-mean and finite
	second moments.
\end{noinds_itemize-2}

\noindent Note that since Assumption \textbf{R2a} only uses current distance estimates, the resulting matrices of barycentric coordinates have constant error biases, i.e.,~$\mathbf{S_P\neq0}$ and $\mathbf{S_B\neq0}$. This is because the relationship of converting distance measurements to barycentric coordinates is non-linear. Clearly, a more accurate scheme, described next, is to utilize the information from past distance measurements, so that one computes the system matrices as a function of the entire past.

\begin{noinds_itemize-2}
	\item \textbf{Assumption R2b: Consistency in distance estimates:} Let $\{Z(k)\}_{k\geq 0}$ be any sequence of inter-node distance 	measurements collected over time. We assume that there exists a sequence of estimates $\{\overline{\mathbf{d}}_{k}\}_{k\geq 0}$ such that, 	for all~$k$, $\overline{\mathbf{d}}_{k}$ can be computed \emph{efficiently} from~$\{Z(s)\}_{s\leq k}$ and we have 
	\begin{equation}
	\label{noisy_dist_ass}
	\mathbb{P}\left[\lim_{k\rightarrow\infty}\overline{\mathbf{d}}_{k}=
	\mathbf{d}^{\ast}\right]=1.
	\end{equation}
	In other words, by a statistically-efficient process (e.g., optimal filtering) based on past distance information, we can
	estimate the required inter-agent distances to arbitrary
	precision as~$k\rightarrow\infty$.
\end{noinds_itemize-2}

We now present two variations of DILOC,\eqref{DILOCeqVect}, namely, \textit{Distributed Localization in Random Environments} (DLRE,~\cite{khan2009distributed}) and \textit{Distributed Localization Algorithm with Noisy Distances} (DILAND,~\cite{khan2010diland}) that improve DILOC under the imperfect scenarios described by Assumptions \textbf{R0}-\textbf{R2}. In particular, DLRE is based on Assumptions \textbf{R0}-\textbf{R2a} and uses only the current distance measurements, while DILAND is based on Assumptions \textbf{R0}, \textbf{R1}, \textbf{R2b}, and utilizes the entire history of distance measurements.

\subsection{DLRE}\label{DLRE}
The DLRE algorithm is described in the following theorem.
\begin{thm}[Theorem 3, \cite{khan2009distributed}]
	\label{thm:drle}
	Under the noise model \textbf{R0}-\textbf{R2a}, the DLRE algorithm given by
	\begin{eqnarray}
	\mathbf{x}_{i}(k+1)&=&\left(1-\alpha\left(k\right)\right)\mathbf{x}_{i}(k)+
	\alpha(k)\nonumber\\
	&&\left[\sum_{j\in\Omega\cap\Theta_{i}}
	\frac{e_{ij}(k)\widehat{p}_{ij}(k)}{q_{ij}}\left(\mathbf{x}_{j}(k)+ \mathbf{v}_{ij}(k)\right)\right.\nonumber\\
	&&\left.+\sum_{j\in\kappa\cap\Theta_{i}}
	\frac{e_{ij}(k)\widehat{b}_{ij}(k)}{q_{ik}}\left(\mathbf{u}_{j}^*
	+\mathbf{v}_{ij}(k)\right)\right],\qquad\label{algass:10}
	\end{eqnarray}
	for~$i\in\Omega$,  with $\alpha(k)\geq0$ satisfying the persistence conditions:
	\begin{eqnarray}\label{alp_perc}
	\alpha(k)\geq 0,\qquad
	\sum_k\alpha(k) = \infty,\qquad
	\sum_k\alpha^2(k) < \infty,
	\end{eqnarray}
	converges almost surely to
	\begin{equation}\label{dlre_convv}
	\lim_{k\rightarrow\infty}\mathbf{X}(k+1) =
	(\mathbf{I-P-S_P})^{-1}(\mathbf{B+S_B})\mathbf{U}(0).
	\end{equation}
\end{thm}

The detailed proof of DLRE can be found in~\cite{khan2009distributed}. Here, we briefly describe the intuition behind the algorithm and its proof. The DLRE update, at time~$k+1$, is essentially a (time-varying) linear  combination of an agent~$i$'s state,~$\mb{x}_i(k)$, and the information that comes from its neighbors, at time~$k$. Self-information from the past is weighted by $1-\alpha(k)$ and since~$\alpha(k)\ra0$, necessary for \eqref{alp_perc} to hold, self-information gets weighted more and more over time and the contribution from  neighbors goes to zero. In fact, because of this dynamic weighting, the iterate~$\mb{x}_i(k)$ must converge as the information from neighbors (that further includes a new noise sample at each iteration) is weighted out over time. However, the rate at which the neighboring contribution is weighted out is carefully selected by the conditions in \eqref{alp_perc}. Loosely speaking, the weighting with~$\alpha(k)$ goes to zero (square summable) but not too fast (infinite sum) allowing the just right amount of information mixing before the neighboring contribution is rejected. The justification behind these arguments come from the stochastic approximation literature, see~\cite{Nevelson} for details.

Next, much the same way as in DILOC, DLRE expresses agent $i$'s location estimate as a function of its non-anchor and anchor neighbors in the triangulation set, i.e.,~$\Omega\cap\Theta_i$ and~$\kappa\cap\Theta_i$, respectively. The communication from these neighboring nodes has the additive noise component, from Assumption~\textbf{R1}. Second, the barycentric coordinate that was $p_{ij}$ in DILOC is replaced by a \textit{barycentric estimate}, $\widehat{p}_{ij}(k)$, that is time-varying and is distorted by a bias and zero-mean error, see Assumption \textbf{R2a}. Finally, random link failures are taken care of by the binary random variable,~$e_{ij}(k)$, in front of the barycentric estimate; this term is further divided by the probability,~$q_{ij}$, of this link being active. Intuitively, division by~$q_{ij}$ ensures that, if a link is active 10\% of the time, the contribution of the corresponding barycentric coordinate is magnified appropriately to counter for the instances the link is dormant. Clearly, DLRE converges to the exact agent locations for unbiased random system matrices, i.e., $\mathbf{S_P=S_B=0}$, as established in Theorem~\ref{thm:drle}. As pointed out earlier, even if the distance estimates are unbiased, the system matrices computed from them may be biased. In such a situation, the DLRE leads to a nonzero steady state error (bias) that is precisely quantified in \eqref{dlre_convv}.

\subsection{DILAND}\label{DILAND}
For the sake of clarity in the following presentation, we analyze DILAND only in the context of noisy distance measurements and assume perfect communication without any link failures and communication noise. The appropriate modifiers in the case of link failures and communication noise are exactly the same as DLRE and hence we ignore them in the following. In order to describe DILAND, we let~$\overline{\mathbf{P}}(\overline{\mathbf{d}}_{k})\triangleq\{\overline{p}_{ij}(k)\}$ and $\overline{\mathbf{B}}(\overline{\mathbf{d}}_{k})\triangleq\{\overline{b}_{ij}(k)\}$ be the matrices of barycentric coordinates computed at time~$k$ from the distance estimate~$\overline{\mathbf{d}}_{k}$, where we will drop the argument in $\overline{\mathbf{P}}$ and $\overline{\mathbf{B}}$ as before. Recall from Assumption \textbf{R2b} that the distance estimate~$\overline{\mathbf{d}}_{k}$ is a function of the entire of history of distance measurements up to time $k$. The DILAND algorithm is described in the following theorem.
\begin{thm}[Theorem 3,~\cite{khan2010diland}]\label{main_DILAND}
	Under the noise model \textbf{R2b}, the DILAND algorithm given by
	\begin{eqnarray}\nonumber
	\mb{x}_i(k+1) &=& (1-\alpha(k))\mb{x}_i(k) + \alpha(k)\\&&
	\left[
	\sum_{j\in\Omega\cap\Theta_i}\overline{p}_{ij}(k)
	\mb{x}_j(k) +
	\sum_{j\in\kappa\cap\Theta_i}\overline{b}_{ij}(k)
	\mb{u}_j^*\right],\qquad\label{diland_it}
	\end{eqnarray}
	$\forall i\in\Omega,$ with~$\alpha(k)\geq0$ satisfying the persistence conditions:
	\begin{eqnarray}\label{alp_perc_dil}
	\alpha(k)\geq 0,\qquad
	\sum_k\alpha(k) = \infty,
	\end{eqnarray} 
converges a.s. to the exact agent locations~$t\rightarrow\infty$.
\end{thm}
The detailed proof of DILAND can be found in~\cite{khan2010diland}. We note that DILAND relies on a key assumption that the distance estimates are a function of all past distance measurements. Due to this, the system matrices~$\overline{\mathbf{P}},\overline{\mathbf{B}}$, at any time~$k$, are a function of past distance measurements, unlike DLRE, where they are assumed to be independent over time. DLRE thus is analyzed in the standard stochastic approximation framework with independent perturbations (see, for example,~\cite{Nevelson}), while the analysis of DILAND is based on stochastic iterations with non-Markovian perturbations. As we mentioned before, DILAND can be extended to cover data packet drops and communication noise but the analysis requires the additional finite summability condition on the weight sequences,~$\alpha(k)$. Finally, we note that an alternative to DILAND can be to estimate inter-node distances by initially collecting and averaging a large number of measurements and then run DLRE. Whether to implement a real-time algorithm that only uses current distance measurements or the entire history, or to run a batch algorithm where distances are estimated a priori, akin to a ``training phase,'' are clear trade-offs and depend on implementation requirements.

DILOC and its extensions to account for environmental imperfections perform localization in static networks. The implementation of DILOC does not immediately carry over to mobile networks, because in mobile networks: 
\begin{inparaenum}[(i)]
	\item agents may move in and out of the convex hull formed by the anchors;
	\item an agent may not be able to find a triangulation set at all times;
	\item the neighborhood at each time step changes as agents move.
\end{inparaenum}
In order to deal with these issues, in the next section, we develop an extension to DILOC that is opportunistic, i.e., the agents update their location estimate when they are successful in finding a valid triangulation set and do not update otherwise. The resulting algorithm is non-deterministic due to arbitrary motion models at the agents and requires a fresh look at the convergence of linear, time-varying systems when their constituent system matrices can take random values. 

\section{Localization in Mobile Networks:\\ Extending Trilateration via Convex Hulls}\label{TRo}
\begin{figure}[!h]
	\centering
	\subfigure{\includegraphics[width=3.5in]{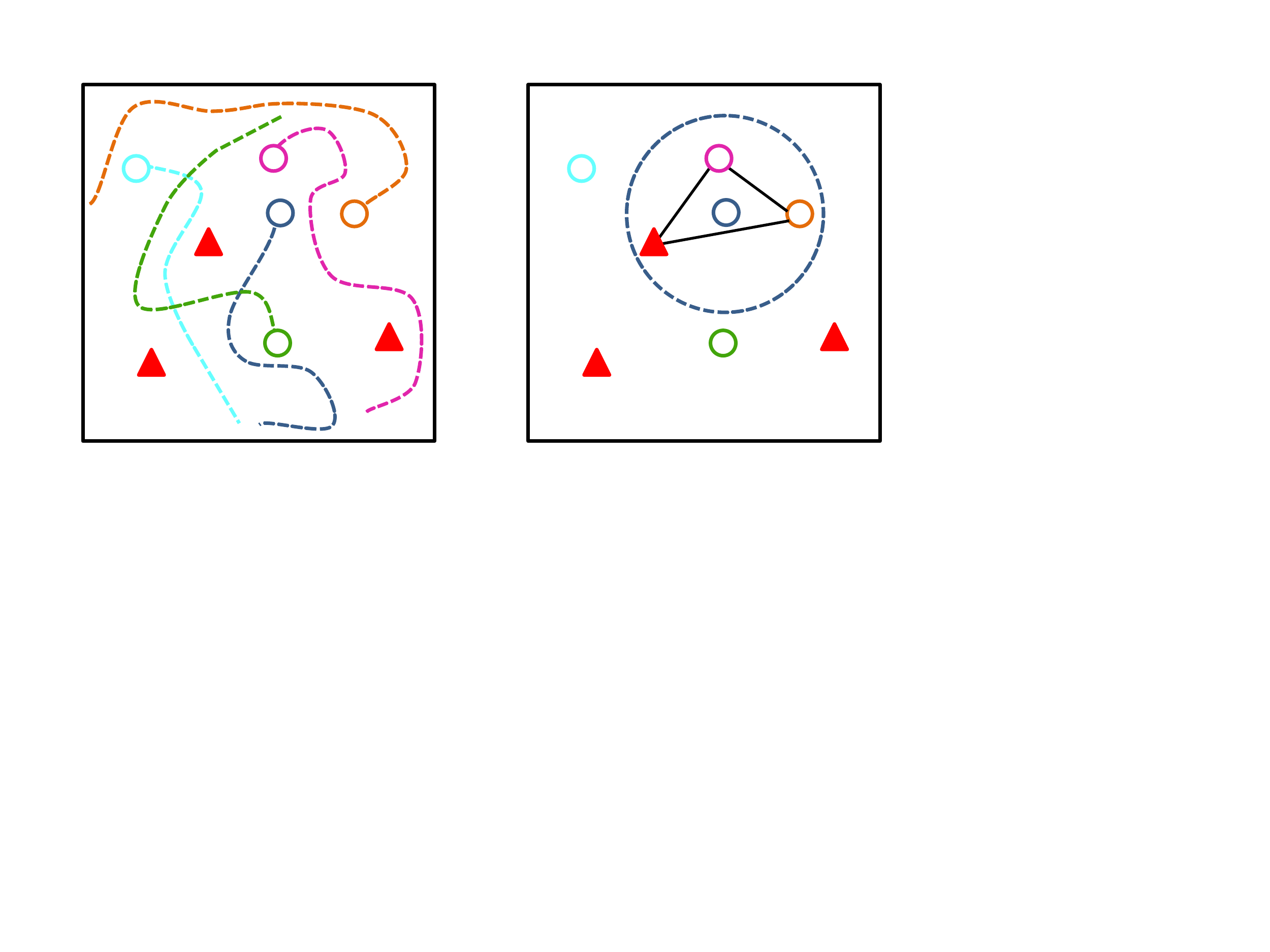}}
	\caption{(Left) Trajectories of the agents according to an arbitrary motion model;
		(Right) Convex-based triangulation; The dashed circle represents the communication range of an agent.}
	\label{f1}
\end{figure}

In this section, we explore the idea of how to extend DILOC to tackle the localization problem in networks of moving objects. Before we proceed with the algorithm, let us explain why the traditional trilateration approach cannot be applied directly for localization of moving objects. Consider a network of $N=5$ agents, with unknown locations, and~$M=3$ anchors, all moving in a bounded region according to an arbitrary motion model, as shown in Fig.~\ref{f1} (Left).	As explained in Section~\ref{BC}, whenever an agent finds $3$ \textit{anchors} in its vicinity, it can localize itself by first measuring its distances to the anchors and then finding the intersection of three circles centered at each anchor location. The problem with this approach is twofold: (i) since each agent has to find at least three neighboring anchors in order to find its location, a relatively large number of anchors has to be deployed in the region of interest; (ii) due to non-linearity of the circle equations, it can not be directly extended to iteratively solve the localization problem. A time-varying extension of DILOC is immediately not applicable as an agent may not find three neighbors at all time.

In order to deal with these issues, we provide an \textit{opportunistic} convexity-based linear algorithm, where an agent updates its location estimate as a convex combination of the states of the neighbors, agents and/or anchors, only if it happens to move inside their convex hull, see Fig.~\ref{f1} (Right). We first consider noiseless scenarios and provide the conditions under which the proposed algorithm converges to the agents' true locations without the presence of any central or local coordinator. We then study the effects of noise on the motion and distance measurements and provide modifications to the original algorithm to counter the undesirable effects of noise. Although our approach is applicable to arbitrary dimensions,
we use~$\mathbb{R}^{2}$ in the remainder of the paper for simplicity and ease of illustration.

\subsection{Distributed Localization Algorithm}\label{sec4}
We now describe the localization algorithm in $\mathbb{R}^2$ while the extension to $\mbb{R}^3$ follows similar arguments. At the beginning (time~$k=0$), each agent is assigned with a random estimate of its initial location. This estimate does not necessarily has to be the coordinates of a point inside the region of motion and is completely random. We then consider two update scenarios for agent~$i$ (in $\Omega$) at any given time~$k>0$:

{\bf{Case (i)}}: If agent $i$ does not find at least $3$ neighbors at time~$k$, i.e., $0\leq\vert{\mathcal{N}}_i(k)\vert < 3$, it does not perform any update except adding the motion vector to its past location estimate:
\begin{eqnarray}\label{18a}
\mb{x}^i_{k+1} = \mb{x}_k^i + \widetilde{\mb{x}}_{k+1}^{i*}.
\end{eqnarray}

{\bf{Case (ii)}}: If agent $i$ finds at least $3$ neighbors at time~$k$, i.e.,~$\vert{\mathcal{N}}_i(k)\vert \geq 3$, it performs the inclusion test, as explained in Section~\ref{BC}, on the possible subsets of $3$ neighbors to determine if it lies inside a convex hull. If there is no subset for which the test is passed, agent~$i$ uses the update in Eq.~\eqref{18a}. If the test is passed and there exists $\Theta_i(k)\subseteq\mc{N}_i(k)$, then it applies the following update:
\begin{eqnarray}\label{18}
\mb{x}^i_{k+1} = \alpha_k\mb{x}_k^i + (1-\alpha_k) \sum_{j\in\Theta_i(k)}a_{ij}^k\mb{x}_k^j + \widetilde{\mb{x}}_{k+1}^{i*},
\end{eqnarray}
in which~$\alpha_k$ is a design parameter such that
\begin{eqnarray}\label{alpk}
\alpha_k=\left\{
\begin{array}{ll}
1, & \forall k~|~\Theta_i(k)=\emptyset,\\
\in\left[\beta,1\right), & \forall k~|~\Theta_i(k)\neq\emptyset,\beta>0.
\end{array}
\right.
\end{eqnarray}
Note that according to \eqref{alpk}, when the agent cannot find a triangulation set, $\alpha_k=1$, the second term in \eqref{18} is zero and the agent's location estimate does not change from time $k$ to $k+1$ (except for incorporating the motion). On the other hand, if the agent finds a triangulation set among its neighbors at time $k$, the weight it assigns to its previous location estimate is lower-bounded by $\beta$. Therefore, \eqref{alpk} guarantees that the agents do not completely forget the valuable location information they have acquired in the past. An agent forgets past information, e.g., by updating with respect to a triangulation set that contains three other agents whose estimates have not been improved from the initial guess; an $\alpha_k$ of at least $\beta>0$ ensures that this does not happen.

By separating the weights assigned to the agents and anchor(s),
we can express the above algorithm in matrix form~as
\begin{eqnarray}\label{eq1}
{\bf{x}}_{{k+1}}={\bf{P}}_{{k}}{{\bf{x}}_{{k}}}+{\bf{B}}_{{k}}{\bf{u}}_{k}+\widetilde{{\bf{x}}}_{{k+1}},\qquad k>0,
\end{eqnarray}
where~${\bf{x}}_{{k}}$ is the vector of agent coordinates at time~$k$,~${\bf{u}}_{{k}}$ is the vector of anchor coordinates at time~$k$, and~$\widetilde{{\bf{x}}}^{*}_{{k+1}}$ is the change in the location of agents at the beginning of the~$k$-th iteration according to the motion model. Also~${\bf{P}}_{{k}}\triangleq\{(\mb{P}_k)_{i,j}\}$ and~${\bf{B}}_{{k}}\triangleq\{(\mb{B}_k)_{i,j}\}$, the system and  input matrices of the above linear time-varying system, contain the weighted barycentric coordinates with respect to the agents with unknown locations, and anchors, respectively.
It can be inferred from~\eqref{alpk} that the self-weight at each agent is always lower bounded, i.e.,
\begin{eqnarray}\label{7}
0<\beta\leq ({\bf{P}}_{k})_{i,i}\leq1, \forall k, i \in \Omega.
\end{eqnarray}

Since anchors play the role of the input in the above linear time-varying system and inject true information into the network, we must also set a lower bound on the weights assigned to the anchor states. To this aim,
we assume that, if an anchor is involved in an update, i.e., for any~$({\bf{B}}_{k})_{i,m} \neq 0$, we have
\begin{eqnarray}\label{8}
0 < \alpha \leq ({\bf{B}}_{k})_{i,m},\qquad \forall k,i\in\Omega, {m\in\Theta_i(k)\cap\kappa}.
\end{eqnarray}
The above assumption implies that, if there is an anchor in the triangulation set it always  contributes a certain amount of information. Since the weights are barycentric coordinates and directly tied to the agent's location inside the convex hull, having a minimum weight on an anchor implies that an update occurs only if an agent lies in an \textit{appropriate} location inside the convex hull of $3$ nodes (in $\mathbb{R}^2$) in the triangulation set (one of which is an anchor). According to~\eqref{8}, the weight assigned to the anchor should be at least~$\alpha$. 

Note that, with the lower bounds on the self-weights and the weights assigned to the anchors according to~\eqref{7} and~\eqref{8}, and assuming, without loss of generality, that at most one update occurs at time~$k$, say at agent~$i$, the system matrix,~${\bf{P}}_{k}$, is either
\begin{enumerate}[(i)]
	\item the \emph{identity}, when there is no update with the neighbors; or,
	
	\item the \emph{identity except a stochastic~$i$-th row}, when no anchor is involved in the update at agent~$i$;
	or,
	
	\item the \emph{identity except a sub-stochastic~$i$-th row}, when there is at least one anchor in the triangulation set of agent~$i$.
\end{enumerate}

We have studied the asymptotic behavior of linear time-varying systems with sub-stochastic system matrices and its applications in~\cite{DBLP:journals/corr/SafaviK14,samCDC,7526779,safavi2015leader}.
In what follows, we use the results on the convergence of an
infinite product of (sub-) stochastic matrices, to provide a sufficient condition for the iterative localization algorithm to converge to the true locations of mobile agents.

\subsection{Distributed localization: Analysis}
We now provide our main result in the following theorem:
\begin{thm}\label{th2}
	Consider a network of $N$ mobile agents moving in a bounded region of interest in the presence of at least one anchor. Assume~\eqref{7} and~\eqref{8} to hold. Then,~\eqref{eq1} asymptotically converges to the true agent locations if each agent establishes a communication path to an anchor, possibly via intermediate agents and over multiple time-steps, in a bounded time, infinitely often.
\end{thm}
Here we provide an intuitive explanation of this theorem. The agent's true locations evolve according to the following updates:
\begin{eqnarray}\label{eq2}
	{\bf{x}}^{*}_{{k+1}}={\bf{P}}_{{k}}{{\bf{x}}^{*}_{{k}}}+{\bf{B}}_{{k}}{\bf{u}}_k+\widetilde{{\bf{x}}}^{*}_{{k+1}}.
\end{eqnarray}
By subtracting~\eqref{eq1} from~\eqref{eq2}, we can find the following error dynamics,
	\begin{eqnarray*}\label{eq3}
	\mb{e}_{k+1}&\triangleq&{\bf{x}}^{*}_{{k+1}}-{\bf{x}}_{{k+1}}={\bf{P}}_{{k}} ({\bf{x}}^{*}_{{k}}-{\bf{x}}_{{k}})={\bf{P}}_{{k}}\mb{e}_{{k}},\nonumber\\
	&=& \prod_{l=0}^{k} {\bf{P}}_{l} \mb{e}_{{0}},
	\end{eqnarray*}
	which converges to zero regardless of the initial error,~$\mb{e}_{{0}}$, if
	\begin{equation}\label{p}
	\lim_{k \rightarrow \infty} \prod_{l=0}^{k} {\bf{P}}_{l}=\mb{0}_{N\times N}.
	\end{equation}
	We have studied the asymptotic behavior of linear time-varying systems with sub-stochastic system matrices in~\cite{DBLP:journals/corr/SafaviK14}. In particular, we show that, by assuming \eqref{7} and \eqref{8} the infinite product of system matrices, $\mb{P}_k$'s, converges to zero if the agents are \textit{connected} to an anchor in a bounded time, infinitely often. By connected, we mean that information from an anchor travels to each agent, possible over several intermediate agents and over multiple time-steps. All that is needed is that this information is received by each agent in a bounded time and hence, each agent receives this information infinitely often over an infinite number of iterations. As a result, \eqref{p} holds and the asymptotic convergence of~\eqref{eq1} to the true agent locations follows.

The reader can refer to~\cite{safavi2017opportunistic} for a detailed proof of the above theorem. In fact, we have proved a much weaker convergence condition than provided in Theorem~\ref{th2}, i.e., the information from an anchor does not necessarily has to reach all agents in a bounded time. In particular, under the assumptions in~\eqref{7} and~\eqref{8}, the infinite product of sub-stochastic system matrices converges to zero as long as the number of iterations, in which the anchor information propagates through the network, grows slower that a certain exponential rate. In other words, the information still propagates in a finite time but the lengths of these information-propagation time-intervals do not need to have a uniform bound. We refer to this criteria as \textit{unbounded connectivity}.

\subsection{How many anchors are required?}
As we discussed in Section~\ref{approaches}, one of the issues with Bayesian approaches for localization is that they require a high density of anchors to achieve accurate results, which in turn adds to the cost of the localization process. Therefore, it is meaningful to investigate the minimal number of anchors that is required for localizing an arbitrary number of moving agents using the localization algorithm proposed in Section~\ref{sec4}. Before providing the main result\footnote{See~\cite{safavi2017opportunistic} for technically rigorous statements and a comprehensive discussion.}, let us denote the motion at agent, $i \in\Omega$, and anchor,~$j \in \kappa$, by~$\mathcal{M}_i$ and $\mathcal{U}_j$, respectively. Suppose agent $1$ is moving along a vertical line; this line forms~$\mathcal{M}_1$ and $\dim \mc{M}_1 =1$. Note that $\mc{M}_i$ (or~$\mathcal{U}_j$) includes all possible locations that the $i$-th agent (or the $j$-th anchor) occupies throughout the localization process.
Now consider another agent, say agent $2$, which is moving along a vertical line parallel to~$\mathcal{M}_1$; in this case we will have
\begin{equation*}
\dim \cup_{i=1,2} \mathcal{M}_i=\dim\mc{M}_2=1.
\end{equation*}
However, if the two lines are linearly independent, they span~$\mathbb{R}^{2}$, and we will have
\begin{equation*}
\dim \cup_{i=1,2} \mathcal{M}_i=2.
\end{equation*}
Note that in the above cases if the agent moves in opposite directions along the line of motion, the dimension of the motion subspace still remains one.

Assuming~$\mathbb{R}^{m}$, in~\cite{safavi2017opportunistic} we show that the motion of the agents and anchors in~$l\leq m$ dimensions allows us to reduce the number of anchors from~$m+1$ by~$l$. The following theorem, provides necessary conditions for the proposed algorithm to track the true location of mobile agents.
\begin{thm}\label{thm3}
	For the linear time-varying dynamics in~\eqref{eq1} to converge to the true locations of the agents, the following conditions must be satisfied:
	\begin{eqnarray}
	\vert \kappa \vert &\geq& 1,\label{33}\\
	\vert \kappa \vert + \vert \Omega \vert &\geq& m+2,\label{34}\\
	\vert \kappa \vert + \dim \underset{i\in\Omega}\cup \mc{M}_i + \dim \underset {j\in\kappa} \cup \mc{U}_j &\geq& m+1 .\label{35}
	\end{eqnarray}
\end{thm}
The proof of the this theorem is beyond the scope of this paper and is treated in~\cite{safavi2017opportunistic}. Since we can provide mobile agents with up to $m$ degrees of freedom in their motion in~$\mathbb{R}^{m}$, according to Theorem~\ref{thm3}, an arbitrary number of mobile agents, $\vert \Omega \vert \geq m+1$, can be localized in the presence of \emph{only one},~$(m+1)-m$, anchor. This is in stark contrast with most Bayesian-based localization algorithms that require a relatively large number of anchors. Also note that the traditional trilateration scheme with static agents requires at least~$3$ anchors in $\mathbb{R}^2$. Therefore, assuming $m+1$ anchors in~$\mathbb{R}^m$ has been standard in all trilateration-based localization algorithms in the literature, see e.g.,~\cite{thomas2005revisiting,navarro1999beacon}.

\subsection{Effects of noise}\label{noise}
So far we assumed that the distance measurements that are used in~\eqref{cmeq} to compute the Cayley-Menger determinants and motion vector in~\eqref{18} are noiseless. In order to take measurement noise and inaccuracy in the motion model into account, we denote the true distance between any two nodes, $i$ and $j$, measured at the time of communication,~$k$, by~$\widetilde{d}^{ij}_k$, and assume that agent~$i$ can only measure a noisy version of this distance:
\begin{equation}
{\widehat{d}_{k}^{ij}} = {\widetilde{d}_{k}^{ij*}} + r^{ij}_k, i\in\Omega,
\end{equation}
where $r_{ij}^k$ is the noise on the distance measurement at time $k$.
We assume that agent~$i$ can obtain a noisy version of its motion:
\begin{equation}
{\widehat{\mb{x}}_{k}^i} = {\widetilde{\mb{x}}_{k}^{i*}} + n_i^k, i\in\Omega,
\end{equation}
in which $n_i^k$ denotes the measurement noise at time $k$.
The noise on the motion and distance measurements degrades the performance of the algorithm and in certain cases the location error is larger than the region of motion. We now provide the following modifications, {\bf{M1-M3}}, to the proposed algorithm to counter the undesirable effects of noise.

{\bf{M1:}} If an agent is located close to the boundary of a convex hull, the noise on the distance measurements may affect the inclusion test results.
According to~\eqref{cmeq2}, in order to get meaningful values for the areas and volumes in~$\mathbb{R}^2$ and $\mathbb{R}^3$, the corresponding Cayley-Menger determinants computed with perfect distance measurements must be negative and positive, respectively. Thus, an agent does not perform an inclusion test if the corresponding Cayley-Menger determinant is positive in $\mathbb{R}^2$, or negative in $\mathbb{R}^3$.

{\bf{M2:}} Even in the case of perfect distance measurements, the inclusion test results may not be accurate due to the noise on the motion, which in turn corresponds to imperfect location updates \textit{at each and every iteration}. To address this issue, even if the inclusion test is passed at time $k$ by a triangulation set, $\Theta_i(k)$, agent $i$ performs an update only if
\begin{eqnarray}\label{iterror}
%i\in\mathcal{C}(\Theta_i(k)),\qquad \mbox{if }
\epsilon_k^{i}=\left\vert\frac{\sum_{j\in\Theta_i(k)}A_{\Theta_i(k)\cup\{i\}\setminus j} - A_{\Theta_i(k)}}{A_{\Theta_i(k)}}\right\vert<\epsilon,
\end{eqnarray}
where~$\epsilon_k^{i}$ is the \textit{relative inclusion test error} at time~$k$ for agent~$i$, and $\epsilon$ is a design parameter.	

{\bf{M3:}} After computing the weights using the noisy distance measurements, the updating agent normalizes the weights assigned to each neighbor in order to preserve convexity.

A detailed analysis addressing noise is beyond the scope of this paper and is discussed in more detail in~\cite{safavi2017opportunistic}. We show the effectiveness of the above modifications via simulation in the next section.

\section{Simulations}\label{simul}
In this section, we provide simulation results to illustrate the effectiveness of the proposed mobile localization algorithm, \eqref{18}, in the presence of one anchor in $\mathbb{R}^2$. Let the true location of the~$i$-th node,~$i\in\Theta$, be decomposed as~${\mb{x}_k^{i\ast}}=[{x_k^{i\ast}}~{y_k^{i\ast}}]$. We consider Random Waypoint motion model,~\cite{perkins2003ad}, which can be expressed for node~$i$ as follows:
\begin{eqnarray*}\label{1}
	{x_{k+1}^{i\ast}}&=& {x_{k}^{i\ast}}+d_{k+1}^i\cos(\theta_{k+1}^i),\\
	{y_{k+1}^{i\ast}}&=& {y_{k}^{i\ast}}+d_{k+1}^i\sin(\theta_{k+1}^i),
\end{eqnarray*}
where~$d^i_{k+1}$ and~$\theta^i_{k+1}$ denote the distance and angle traveled by node~$i$, between time~$k$ and~$k+1$, and are random. We choose,~$d^i_{k}$ and~$\theta^i_{k}$, to have uniform distribution over the intervals~$[0~d_{\max}=5 \mbox{m}]$ and~$[0~2\pi]$, respectively, such that the agents remain inside the region of interest. We assume that the noise on the distance and angle that agent~$i$ travels at time $k$ are Gaussian with zero mean and the following variances,
\begin{eqnarray*}
	{\sigma_d^i}^2&=&{K_d}^2D^i_k,\\
	{\sigma_{\theta}^i}^2&=&{{K_{\theta}}^2} D^i_k,
\end{eqnarray*}
where~$D^i_k$ represents the total distance that agent $i$ has traveled up to time $k$. We also assume that the noise on the distance measurement at time $k$ is normal with zero mean and the variance~of~${\sigma_r^i}^2={K_r}^2 k$. Thus, the variances of the odometry measurements are proportional to the total distance an agent has traveled, and the variance on the distance measurements (to the neighboring agents) increases with time. Such assumptions are standard, e.g., in the robotics literature~\cite{martinelli2005observability,chong1997accurate}.

\begin{figure}[!h]
	\centering
	\subfigure{\includegraphics[width=1.72in,height=1.72in]{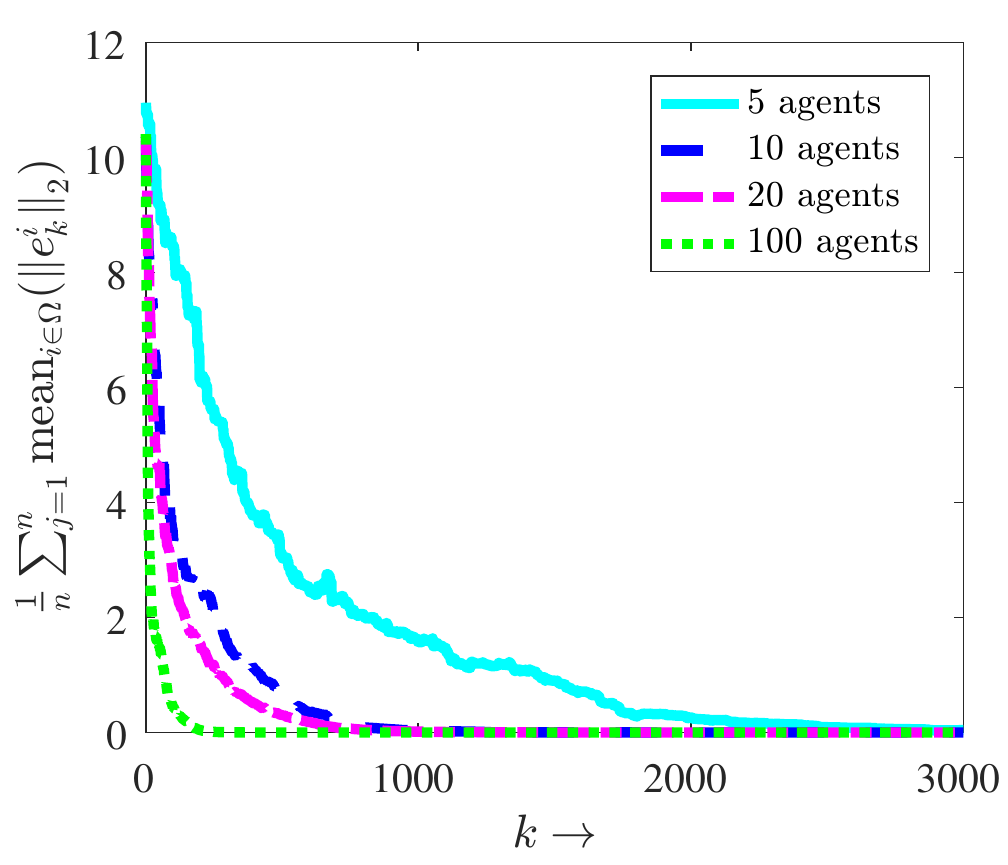}}
	\subfigure{\includegraphics[width=1.72in,height=1.72in]{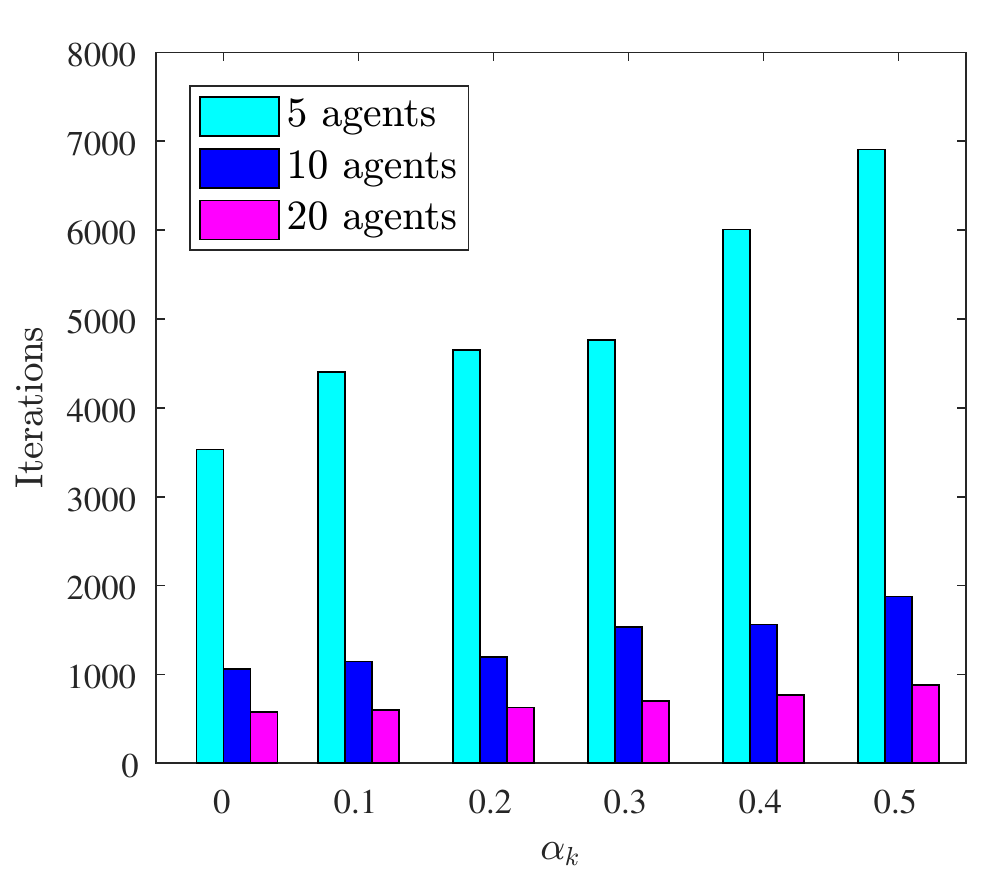}}
	\caption{(Left) Convergence;
		(Right) Effect of self-weights.}
	\label{f2}
\end{figure}
\begin{figure}[!h]
	\centering
	\subfigure{\includegraphics[width=3.65in]{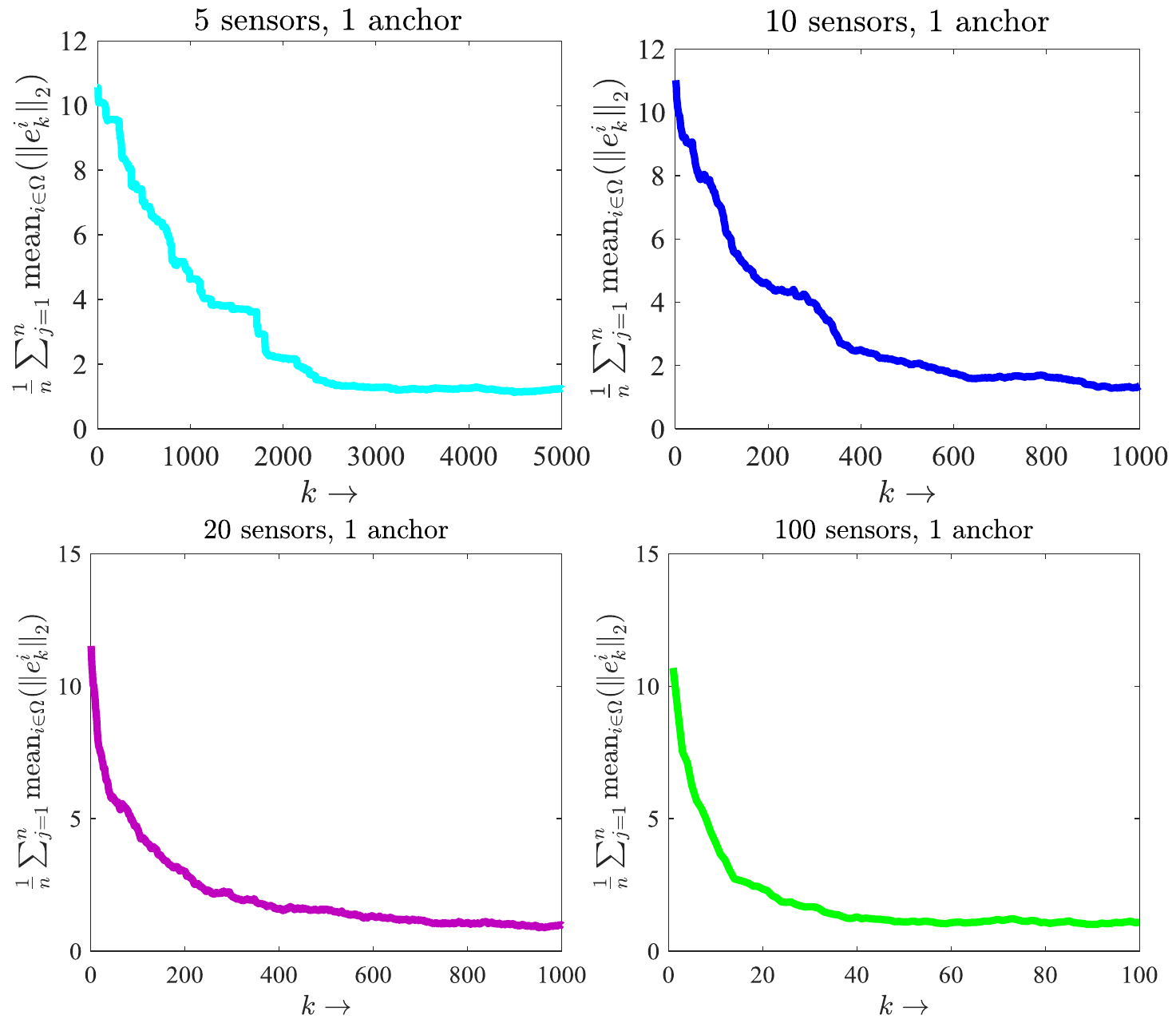}}
	\caption{Modified algorithm with noise.}
	\label{f3}
\end{figure}

In the beginning, all nodes are randomly deployed within a region, which is a $20~\mbox{m} \times 20~\mbox{m}$ square. We set the communication radius to $r=2~\mbox{m}$. All agents are initially assigned with random location estimates. We set~$\alpha_k=\beta=0.01$ to ensure that the agents do not completely forget the past information and~$\alpha=0.01$ to guarantee a minimum contribution from the anchor when it is involved in an update. To characterize the convergence, we choose the second norm of the error vector,~$\mb{e}_k$. It can be seen in Fig.~\ref{f2} (Left) that the error in the localization algorithm converges to zero in the absence of noise for networks with \textit{one mobile anchor} and $5$,~$10$,~$20$, and~$100$ mobile agents.
Fig.~\ref{f2} (Right) shows the effect of increasing the self-weight on the rate of convergence for networks with $5$,~$10$,~$20$ mobile agents.
\begin{figure}[!h]
	\centering
	\subfigure{\includegraphics[width=1.72in,height=1.72in]{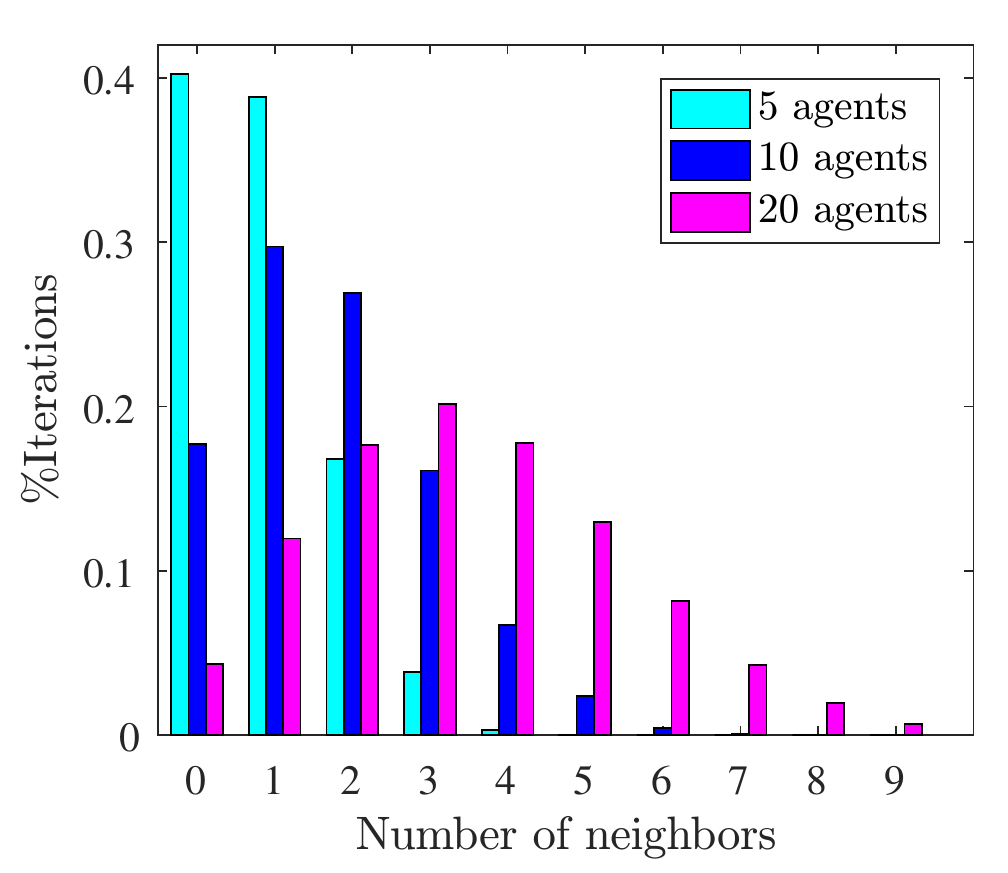}}
	\subfigure{\includegraphics[width=1.72in,height=1.72in]{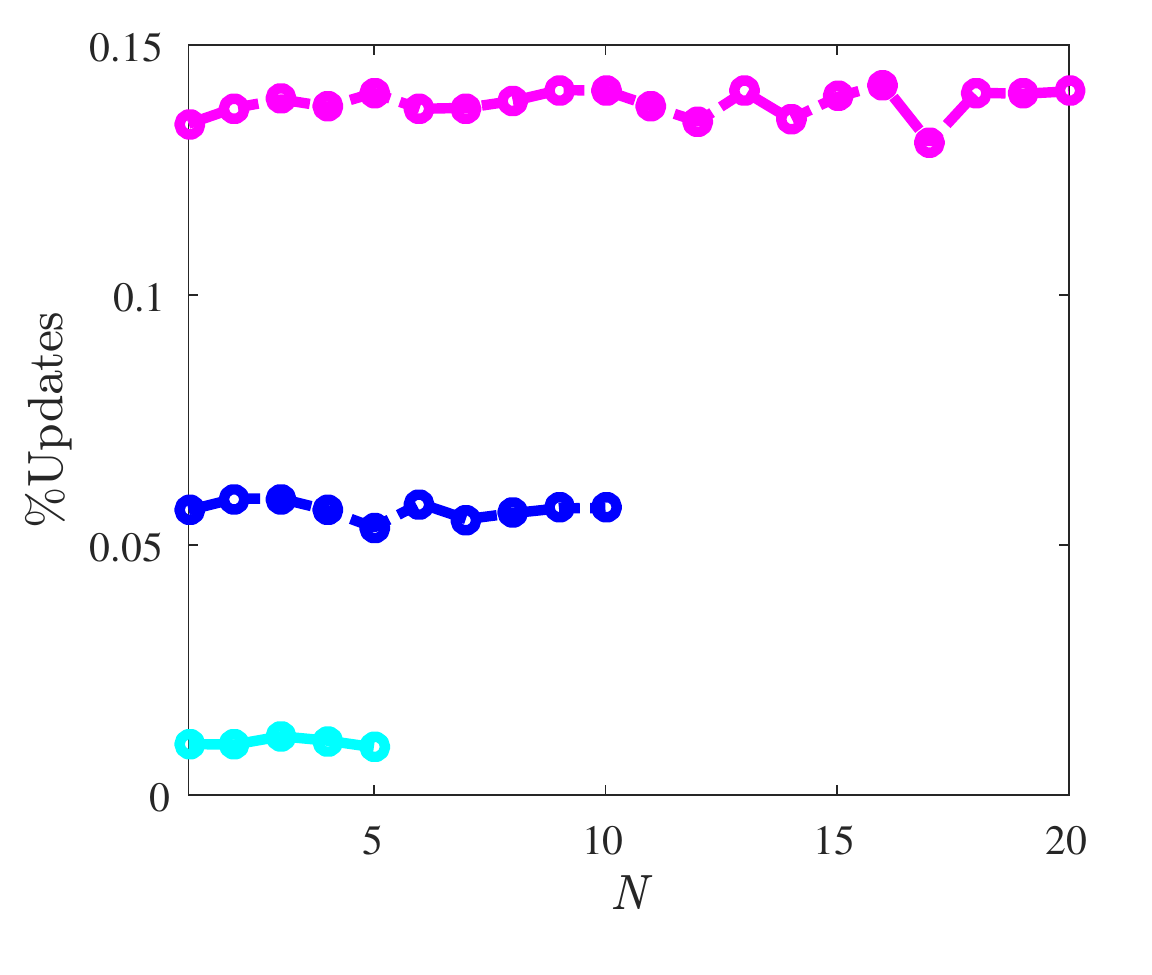}}
	\caption{(Left) Number of neighbors;
		(Right) Number of updates/iterations.}
	\label{f4}
\end{figure}
In Fig.~\ref{f4} (Left), we show the percentage of the iterations that an agent finds a different number of neighbors in those networks. For example, in a network of $5$ agents and one anchor, an agent finds no neighbors during $40\%$ of the iterations and finds only one neighbor in $38\%$ of the iterations. Fig.~\ref{f4} (Right) shows the ratio of the total number of updates to the number of iterations. In our simulations, we fix the number of iterations to $3000$, and take the average over $n=20$ Monte Carlo simulations. On average, an agent in networks with one anchor and $5$, $10$, and $20$ agents, performs~$31$, $171$, and $422$ updates, respectively.
Finally, Fig.~\ref{f3} shows that with ${K_d}=K_{\theta}=K_r=5*10^{-3}$, by modifying the algorithm according to Section~\ref{noise} (with $\epsilon=20\%$), the localization error is bounded by the communication radius.

\section{Conclusions}\label{conc}
Motivated by the key role that location-awareness is expected to play in Fifth generation~(5G) networks, as well as the rapidly growing interest on Internet-of-Things (IoT) and the significance of location information on a variety of IoT applications such as healthcare and surveillance, this paper considers localization in dynamic networks of mobile \emph{objects}. We review Bayesian-based solutions as currently the most commonly used approaches to localization, especially in mobile networks; these solutions are non-linear, mainly centralized, and require a high density of anchors, nodes with known locations, in order to achieve accurate results. In order to avoid high computation costs and communication overhead associated with Bayesian-based solutions, we describe DILOC, an alternative \textit{linear} framework for localization in static networks, which only requires local measurements, local communication, and low-order computation at each agent. Linearity is obtained by reparametrization of the agents' location through barycentric coordinates. These are based on local neighborhood geometry and may be computed by Cayley-Menger determinants. The Cayley-Menger determinants involve only local inter-agent distances. After studying the convergence of DILOC and its robustness to noise, we extend it to mobile scenarios, suitable for 5G enabled IoT environments, in which agents, users, and (possibly) anchors are dynamic. We show that the algorithm in Section~\ref{TRo} can localize an arbitrary number of mobile agents in the presence of \textit{at least one anchor}. Section~\ref{TRo} also considers the effects of noise on the localization algorithm for mobile agents. 

\bibliographystyle{IEEEbib}
\bibliography{bibliography}
\end{document}